\documentclass[%
reprint,
superscriptaddress,
amsmath,amssymb,
aps,
prb,
]{revtex4-2}
\usepackage{lipsum}
\usepackage{lineno}
\usepackage[dvipsnames]{xcolor}
\usepackage{graphicx}
\usepackage[caption=false]{subfig}
\usepackage{dcolumn}
\usepackage{bm}
\usepackage{hyperref}
\usepackage{xr}
\hypersetup{
  colorlinks,
  citecolor=blue,
  linkcolor=blue,
  urlcolor=blue}
\usepackage{color}
\usepackage{booktabs}

\newif\ifshowcomments
\showcommentstrue

\usepackage{xcolor}

\newcommand{\RNum}[1]{\uppercase\expandafter{\romannumeral #1\relax}}

\newcommand{\beginsupplement}{
        \setcounter{table}{0}
        \renewcommand{\thetable}{S\arabic{table}}
        \setcounter{figure}{0}
        \renewcommand{\thefigure}{S\arabic{figure}}
        \setcounter{equation}{0}
        \renewcommand{\theequation}{S\arabic{equation}}
        \setcounter{page}{1} 
        \renewcommand{\thepage}{S\arabic{page}} 
        \onecolumngrid
}
\begin{document}
\preprint{APS/123-QED}
\title{Benchmarking quantum simulation with neutron-scattering experiments}

\author{Yi-Ting Lee}
\thanks{These authors contributed equally to this work.}
\affiliation{Department of Materials Science and Engineering, University of Illinois at Urbana-Champaign, Urbana, Illinois 61801, USA}

\author{Keerthi Kumaran}
\thanks{These authors contributed equally to this work.}
\affiliation{Department of Physics and Astronomy, Purdue University, West Lafayette, Indiana 47907, USA}
\affiliation{Quantum Science Center, Oak Ridge National Laboratory, Oak Ridge, Tennessee 37831, USA}

\author{Bibek~Pokharel}
\altaffiliation{Corresponding author}
\affiliation{Quantum Science Center, Oak Ridge National Laboratory, Oak Ridge, Tennessee 37831, USA}
\affiliation{IBM Quantum, IBM Thomas J. Watson Research Center, Yorktown Heights, New Tork, USA}

\author{Allen Scheie}
\affiliation{MPA-Q, Los Alamos National Laboratory, Los Alamos, New Mexico 87545, USA}

\author{Colin L. Sarkis}
\affiliation{Neutron Scattering Division, Oak Ridge National Laboratory, Oak Ridge, Tennessee 37831, USA}

\author{Stephen E. Nagler}
\affiliation{Neutron Scattering Division, Oak Ridge National Laboratory, Oak Ridge, Tennessee 37831, USA}

\author{D.~Alan~Tennant}
\affiliation{Department of Physics and Astronomy, University of Tennessee, Knoxville, Tennessee 37996, USA}
\affiliation{Department of Materials Science and Engineering, University of Tennessee, Knoxville, Tennessee 37996, USA}

\author{Travis S.~Humble}
\affiliation{Quantum Science Center, Oak Ridge National Laboratory, Oak Ridge, Tennessee 37831, USA}

\author{Andr\'e Schleife}
\affiliation{Department of Materials Science and Engineering, University of Illinois at Urbana-Champaign, Urbana, Illinois 61801, USA}
\affiliation{Materials Research Laboratory, University of Illinois at Urbana-Champaign, Urbana, Illinois 61801, USA}

\author{Abhinav Kandala}
\affiliation{Quantum Science Center, Oak Ridge National Laboratory, Oak Ridge, Tennessee 37831, USA}
\affiliation{IBM Quantum, IBM Thomas J. Watson Research Center, Yorktown Heights, New Tork, USA}

\author{Arnab Banerjee}
\altaffiliation{Corresponding author}
\affiliation{Department of Physics and Astronomy, Purdue University, West Lafayette, Indiana 47907, USA}
\affiliation{Quantum Science Center, Oak Ridge National Laboratory, Oak Ridge, Tennessee 37831, USA}

\begin{abstract}
Realistic simulation of quantum materials is a central goal of quantum computation. Although quantum processors have advanced rapidly in scale and fidelity, it has remained unclear whether pre–fault-tolerant devices can perform quantitatively reliable material simulations. 
We demonstrate that a superconducting quantum processor operating on up to 50 qubits can already produce meaningful, quantitative comparisons with inelastic neutron-scattering measurements of KCuF$_3$, a canonical realization of a gapless Luttinger liquid system with a strongly correlated ground state and a spectrum of emergent spinons.
The quantum simulation is enabled by a quantum–classical workflow for computing dynamical structure factors (DSFs).
The resulting spectra are benchmarked against experimental measurements using multiple metrics, highlighting the impact of circuit depth and circuit fidelity on simulation accuracy.
Finally, we extend our simulations to a 1D XXZ Heisenberg model with next-nearest-neighbor (NNN) interactions and a strong anisotropy, producing a gapped excitation spectrum, which could be used to describe the CsCoX$_3$ compounds above the N\'eel temperature.
Our results establish a framework for computing DSFs for quantum materials in classically challenging regimes of strong entanglement and long-range interactions, enabling quantum simulations that are directly testable against laboratory measurements.
\end{abstract}

\maketitle


Accurately predicting properties of quantum materials has been enabled by decades of advances in classical computational methods~\cite{foulkes2001quantum,schollwock2011density,orus2019tensor,saad2011numerical}.
However, strongly correlated systems with long-range entanglement and complex real-time dynamics remain beyond the reach of these approaches.
Quantum computers offer a potential alternative to addressing this challenge~\cite{alexeev2024quantum,clinton2024towards}, as envisioned by Feynman~\cite{feynman2018simulating}.
Rapid advances in qubit number and gate fidelity have enabled studies of both static~\cite{robledo2025chemistry,yu2025quantum} and dynamical~\cite{haghshenas2025digital, google2025observation,farrell2025digitalquantumsimulationsscattering} properties of many-body systems at scales beyond exact diagonalization.
Yet, despite recent advances in algorithms and hardware, resource estimates suggest that extending these simulations to realistic materials is expected to require circuit depths and error rates that are beyond near-term capabilities~\cite{clinton2024towards}.
It has therefore been unclear whether current, pre-fault tolerant quantum computers can ever perform quantitatively reliable many-body simulations of quantum materials that can be closely compared with laboratory measurements.

Inelastic neutron scattering (INS) provides an experimentally accessible pathway to bridge quantum simulation with laboratory measurements. 
In an INS experiment, incident neutrons scatter off magnetic ions in a material, exchanging both energy and momentum with its spins. 
The measured signal can be used to compute two-point dynamical correlation functions that capture how local excitations propagate through the system in space and time. 
Expressed in momentum and frequency space, these correlations form the dynamical structure factor (DSF), which directly reveals the spectrum of collective excitations. 
Because many quantum materials can be modeled by spin-1/2 Hamiltonians that are naturally mapped onto qubit systems, digital quantum processors can be used to compute the DSF~\cite{chiesa2019quantum,baez2020dynamical,eassa2024high,bauer2025progress, related_work_1, related_work_2, related_work_3}.
This correspondence provides a rigorous framework for validating quantum simulation against experimental observables.

Classically, DSFs can be accessed using approaches such as semiclassical spin-wave theory~\cite{fishman2018spin}, exact diagonalization~\cite{Liu2019ED}, Krylov-space methods~\cite{Luitz2019Krylov}, and tensor-network techniques including t-DMRG~\cite{Fratus2016tDMRG}. While powerful, these methods are ultimately limited by Hilbert-space growth or entanglement proliferation during real-time evolution, particularly in strongly interacting or quantum-critical regimes.
Indeed, computing DSFs for general local Hamiltonians is believed to inherit the BQP-hardness of real-time quantum dynamics~\cite{baez2020dynamical}, highlighting the potential advantage of quantum processors for accessing these observables.


Here we demonstrate that pre-fault-tolerant, programmable quantum processors can serve as reliable tools for simulating experimentally measurable spectra of quantum materials.
We perform a direct experimental validation by benchmarking quantum simulations against INS measurements of the quasi-1D antiferromagnet KCuF$_3$, whose spectrum is a canonical realization of a Tomonaga–Luttinger liquid exhibiting fractionalized spinon excitations~\cite{lake2005quantum,scheie2022quantum}.
Using 50-qubit circuits executed on superconducting quantum processors, we compute spatiotemporal correlation functions and reconstruct the DSF for quantitative comparison with experiment. 
We show that the agreement between simulation and experiment is enabled by reductions in two-qubit error rates towards 0.1\%, and the intrinsic robustness of the DSF estimation to noise-induced damping of the correlation functions. 
We further extend this framework to a non-integrable XXZ model with next-nearest-neighbor (NNN) interactions, relevant to CsCoX$_3$ (X = Halides) above their Néel temperature~\cite{matsubara1991magnetic,matsubara1991two,shiba2003exchange}, demonstrating applicability beyond analytically solvable regimes. 
These results establish a pathway for quantum simulations that are directly testable against laboratory measurements, providing a benchmark for the scientific capabilities of near-term quantum processors.

\begin{figure}
\centering
\includegraphics[width=0.48\textwidth]{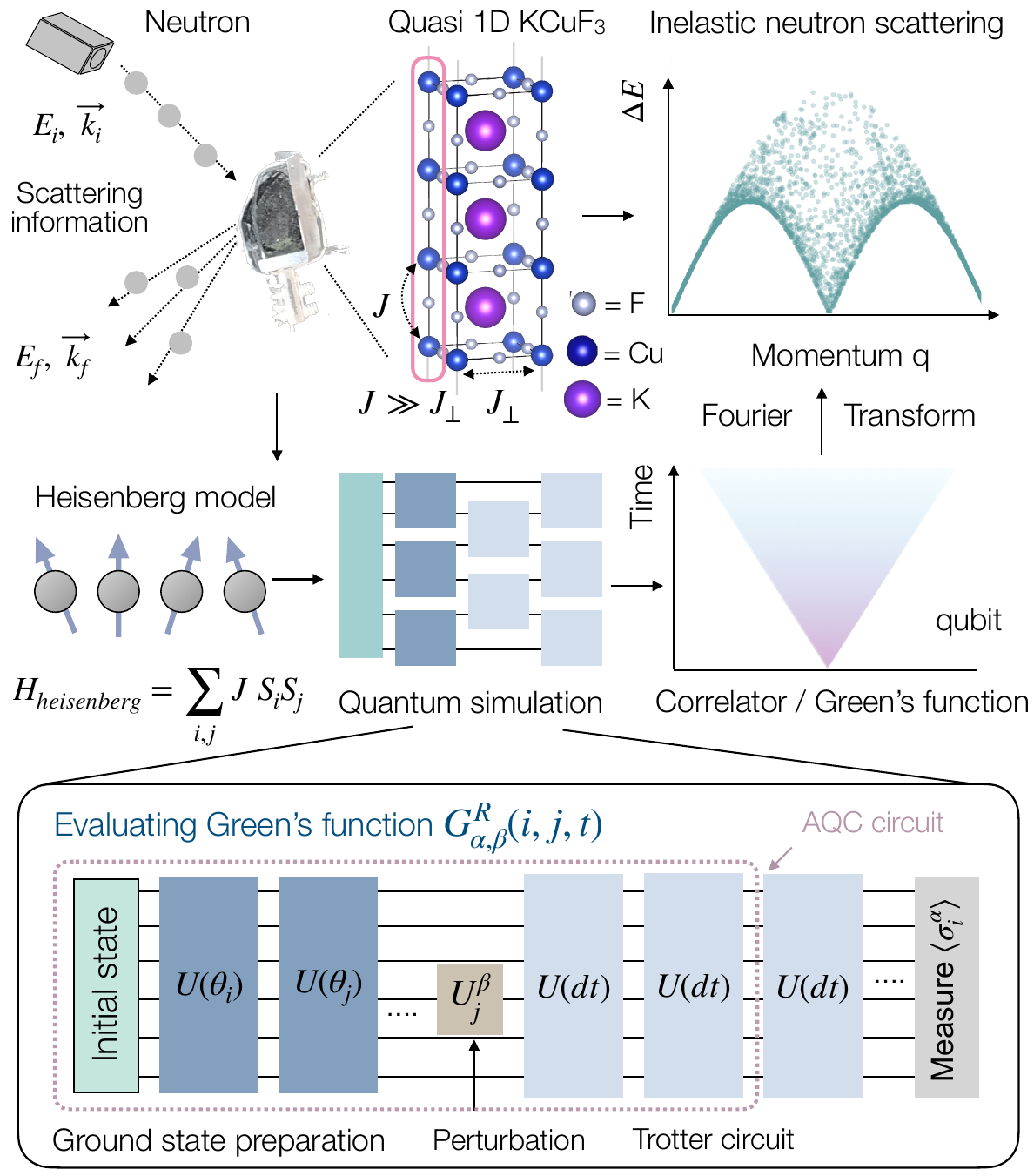} 
\caption{\textbf{Obtaining spectra via INS and quantum simulation.}
Protocol for computing the DSF and comparing with INS data (Using KCuF$_3$ as an example).
Neutron scattering measures the excitation spectrum in momentum space. Using a model spin Hamiltonian, we compute the retarded Green’s function (RGF) and Fourier transform it to obtain the DSF. The top left panel shows the single crystal of KCuF$_3$ (6.86 g) used to obtain INS data.
In the quantum simulation, the ground state is prepared variationally~\cite{yu2023simulating}, a local perturbation is applied, the system is evolved, and $\sigma_i^{\alpha}$ is measured to extract the RGF.
}
\label{overview}
\end{figure}

This quantitative comparison is grounded in the formal mapping of the simulated DSF $S(q, \omega)$ to the experimentally accessible cross-section. In an INS experiment, thermal and cold neutrons from reactor or spallation sources serve as a non-destructive and highly sensitive probe of spin dynamics at the atomic scale, yielding a signal that corresponds to the magnetic neutron scattering cross-section~\cite{lovesey1984theory},
\begin{equation}
\label{eq:INS}
I({\bf q},\omega) = (\gamma r_0)^2 \frac{k_f}{k_i} \left| \frac{g}{2} F({\bf q}) \right|^2
\sum_{\alpha,\beta = x,y,z}
(\delta_{\alpha\beta} - \hat{q}_\alpha \hat{q}_\beta)\,
S^{\alpha,\beta}({\bf q},\omega).
\end{equation}
 In this expression, \( {\bf q} \) denotes the momentum transfer, \( k_i \) and \( k_f \) are the wave vectors of the incident and scattered neutrons, respectively, \( r_0 \) is the electron (spin) radius, $\gamma$ is the neutron gyromagnetic ratio, \( g \) is the Lande' g-factor, and \( F({\bf q}) \) is the magnetic form factor, which depends on the isotope. 
Importantly, the tensor \( S_{\alpha,\beta}({\bf q},\omega) \) encodes the spin–spin dynamical correlations of the system and fully characterizes its magnetic excitation spectrum.

Within the framework of the fluctuation-dissipation theorem~\cite{kubo1966fluctuation}, assuming the absence of any temperature-driven phase transitions, the DSF at experimentally relevant temperatures can be obtained directly from the imaginary part of the retarded Green’s function (RGF) \( G^{R}_{\alpha,\beta}({\bf q},\omega) \):
\begin{equation}
\label{eq:DSF_GF}
S_{\alpha,\beta}({\bf q},\omega)
=
-\frac{1}{\pi}\,[1+n_B(\omega)]\,
\mathrm{Im}\!\left[G^{R}_{\alpha,\beta}({\bf q},\omega)\right],
\end{equation}
where \( n_B(\omega) \) is the Bose–Einstein distribution function, which vanishes at zero temperature.
Crucially, the RGF can be evaluated efficiently in the position–time domain, \( G^{R}_{\alpha,\beta}(i,j,t) \), using quantum simulation protocols based on local unitary perturbations applied at individual lattice sites~\cite{baez2020dynamical}.
These protocols enable the direct measurement of spatiotemporal correlation functions on quantum hardware, which are then Fourier transformed to obtain the DSF in momentum and frequency space, as shown in Fig.~\ref{overview} and detailed in Supplementary Section~\ref{quantum_implementation_details}. 
Overall, the DSF is experimentally accessible both through INS experiment on a real material and via digital quantum simulation on a programmable quantum computer.

\begin{figure*}
\centering
\includegraphics[width=0.98\textwidth]{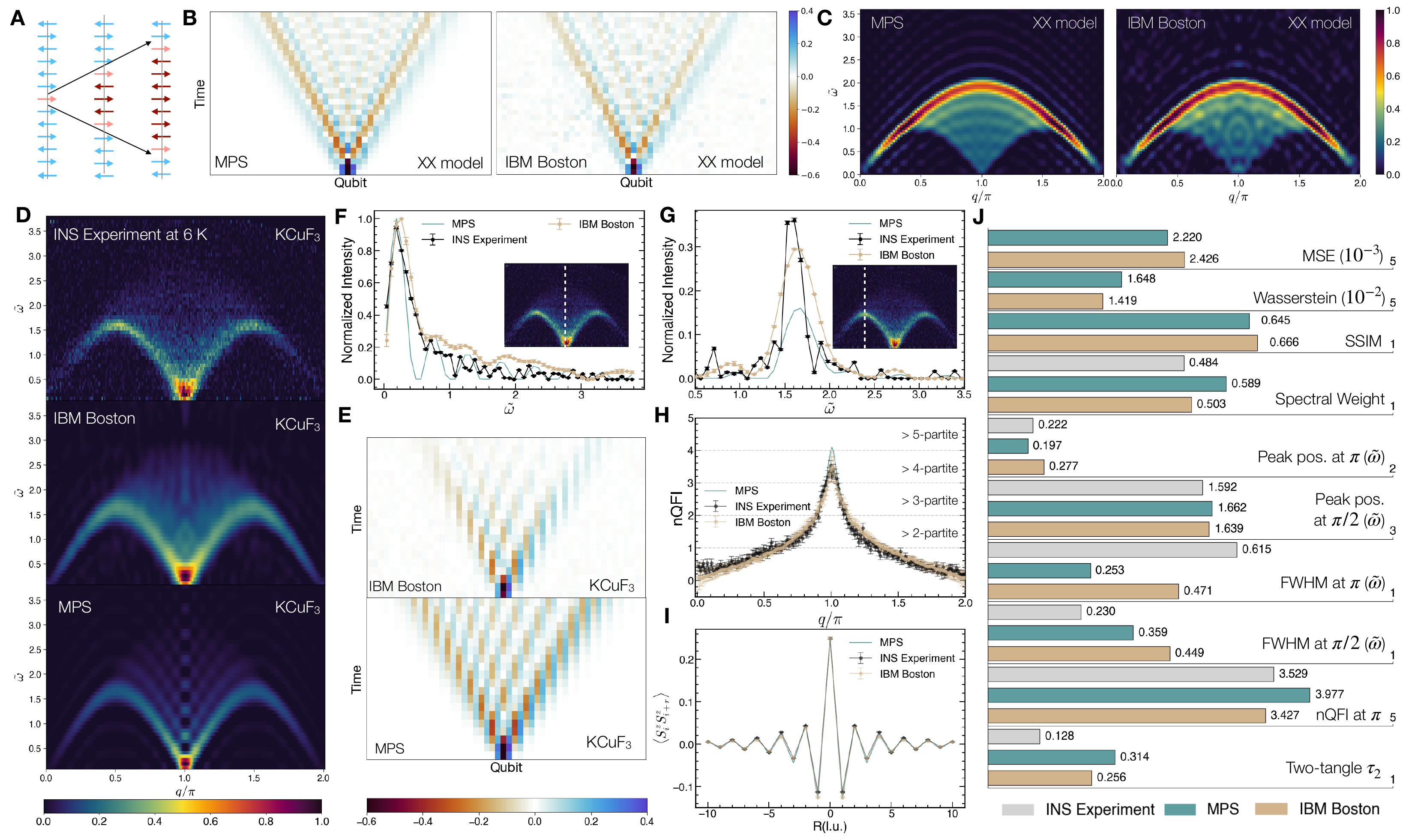}
\caption{ \label{KCuF3_spectrum}
\textbf{INS spectra for the XX model and KCuF$_3$.}
(\textbf{A}) Schematic of the spread of fractionalized excitation (spinon) following neutron scattering.
(\textbf{B},\textbf{C}) Retarded Green’s function (RGF) and DSF from 50-qubit MPS and quantum simulations of the XX model.
(\textbf{D},\textbf{E}) Corresponding RGF and DSF for KCuF$_3$, compared with experimental INS data obtained using the MAPS spectrometer at ISIS Facility from Ref.~\cite{lake2013multispinon}.
The RGF exhibits a light-cone structure, while finite-size, finite-time effects and hardware noise lead to pixelation and spectral broadening.
(\textbf{F},\textbf{G}) Line scans of the KCuF$_3$ spectrum at $q=\pi$ and $q=\pi/2$.
(\textbf{H},\textbf{I}) Quantum Fisher information and spin-spin correlations.
(\textbf{J}) Quantitative benchmarking of experimental, MPS, and quantum results using image-based metrics, spectral weight, line scans, and entanglement witnesses.}
\end{figure*}

\subsection*{Quantum simulation of INS spectrum}
We first focus on evaluating the INS spectrum of KCuF$_3$, which is a prototypical quasi-1D antiferromagnet whose magnetic properties have been extensively characterized by INS experiments.
While KCuF$_3$ crystallizes in a tetragonal structure, orbital ordering leads to strong antiferromagnetic superexchange interactions along the $c$ axis and much weaker couplings perpendicular to it~\cite{pavarini2008mechanism}, resulting in an emergent 1D quantum magnet.
Both experiment~\cite{lake2013multispinon} and theory indicate that KCuF$_3$ is well described by the one-dimensional spin-$1/2$ Heisenberg model, which corresponds to the isotropic point ($\epsilon=1$) of the integrable 1D XXZ Hamiltonian:
\begin{equation}
\label{NN_model}
    H_{\text{NN}} = 2J \sum_{i=1}^{N-1}
    \left(S_i^Z S_{i+1}^Z + 
    \epsilon (S_i^X S_{i+1}^X +
    S_i^Y S_{i+1}^Y )
    \right),
\end{equation}
with exchange interaction $J$ and anisotropy parameter $\epsilon$.
This isotropic model has a storied history, beginning with Bethe’s 1931 ansatz for the eigenstates~\cite{bethe1931theorie}, followed by analytical results for low-lying excitations~\cite{des1962spin} and the exact DSF in the thermodynamic limit~\cite{caux2006four}.
It is also a paradigmatic strongly correlated, quantum-critical system~\cite{lake2005quantum} exhibiting fractionalized excitations~\cite{bethe1931theorie,faddeev1981spin} and superdiffusive spin transport, with correlations decaying as $t^{-2/3}$~\cite{scheie2021detection,integrable_9}. 
As a result, it serves as a trusted benchmark for prior quantum simulation studies~\cite{Keenan_2023,Elliot,kumaran2025superdiffusionresilienceheisenbergchains,lee2026digital}.


Our quantum-computing workflow (Fig.\ref{overview}) closely mimics the local $S = 1$ spin-flip excitation natively generated during an inelastic neutron scattering event. In 1D XXZ chains, this excitation fractionalizes into two spinons (Fig.~\ref{KCuF3_spectrum}A).
These spin-$1/2$ quasiparticles propagate independently, producing a broad continuum — an unambiguous signature of fractionalized dynamics in contrast to the conventional sharp magnon modes characteristic of bosonic spin waves.

To clarify the fundamental spectral changes across different transport regimes, we also consider the Hamiltonian in XX limit of Eq.~\ref{NN_model} where the longitudinal ZZ interaction vanishes. This regime exhibits ballistic transport with $t^{-1}$ decay.
For the XX model, the dynamics manifest as a sharp light-cone for the retarded Green’s function (RGF) (Fig.~\ref{KCuF3_spectrum}B)~\cite{scheie2022quantum} and a DSF with spectral weight concentrated near the upper boundary of the continuum (Fig.~\ref{KCuF3_spectrum}C), consistent with theoretical expectations of ballistic spinon
propagation in the XX limit.
Close qualitative agreement with this solvable reference point establishes a clear baseline against known transport behavior.

At the isotropic point, relevant to KCuF$_3$, the two-spinon continuum forms a gapless, dispersive lower boundary (Fig.~\ref{KCuF3_spectrum}D), reflecting the emergence of slow spinons and dynamics reminiscent of a 1D quantum spin liquid. 
While weak interchain coupling in KCuF$_3$ leads to an ordered ground state, and consequently the lowest energy spectrum at 6 K reflects a subtle dimensional crossover exhibiting magnons and a longitudinal “Higgs” mode \cite{lake_longitudinal,lake2005quantum}, the 1D model remains the essential framework for capturing the core fractionalized dynamics. 
Even at low temperatures, the fractional excitations dominate the measured response function over most of the energy range.
The RGF in this model exhibits antiferromagnetic correlations pinned to the light-cone boundary together with persistent oscillations inside the cone (Fig.~\ref{KCuF3_spectrum}E).
This long-lived oscillatory state, termed the \emph{quantum wake}~\cite{scheie2022quantum}, is heuristically understood as interference between counterpropagating spinons and antispinons.

The redistribution of spectral weight between the upper and lower boundaries of the continuum thus provides a direct fingerprint of the underlying transport regime.
Enhanced intensity near the lower boundary reflects slower excitation dynamics and the collapse of ballistic modes from the upper boundary.
Taken together, the evolution of spectral weight and the RGF spreading pattern can offer a diagnostic for distinguishing superdiffusive from ballistic transport.

The quantum simulations faithfully reproduce the key qualitative spectral features, including the gapless continuum and signatures of spin fractionalization.  
Despite the strong suppression of long-time signal amplitudes in the RGF (Fig.~\ref{KCuF3_spectrum}E) in deep circuits, which generates spurious signal and broadens the spectrum, the resolution remains sufficient to enable qualitative comparison, owing to the robustness of the Fourier transform against signal damping.
Notably, this smearing mimics finite-temperature or finite-lifetime broadening, but it does not originate from intrinsic physics and is expected to diminish systematically as quantum devices improve.

\subsection*{Benchmarking}

To quantitatively assess the agreement between quantum simulations and INS spectra, we first employ three complementary similarity metrics: mean squared error (MSE), Wasserstein distance, and the structural similarity index measure (SSIM).
While MSE captures the average local pixel-wise deviation and the Wasserstein distance captures global similarity, SSIM provides a structure-aware comparison, with values ranging from 1 (perfect similarity) to 0 (no similarity) to -1 (perfect antisimilarity). 
The MPS results, computed using identical system size, time discretization, and Fourier procedures, set an upper bound on the accuracy achievable in the absence of hardware noise. Overall, all three metrics confirm that the quantum simulations reproduce the INS spectra with fidelity comparable to MPS benchmarks (Fig.~\ref{KCuF3_spectrum}J).

However, as these generic metrics are not physics-informed, we further analyze spectral properties that are directly relevant to condensed-matter experiments: spectral weight, peak positions, linewidths, and entanglement properties. In terms of spectral weight, the main experimental dispersion carries approximately 48\% of the total intensity. MPS slightly overestimates this value (58.9\%), the $ibm\_boston$ results yield values closer to experiment ($\sim$50\%). This agreement reflects the presence of hardware noise that smears the intensity and lowers the observable spectral weight toward experimental levels.

To align our validation with standard experimental diagnostics, we now turn to the spectral lineshapes and peak positions through a line-scan analysis (Figs.~\ref{KCuF3_spectrum}F,G).
The uncertainty in the estimated peak position at the $\pi$ point remains below 10\% of the full width at half maximum (FWHM) for both MPS and quantum simulations, increasing to 20–30\% at $q=\pi/2$.
At low frequencies, $q=\pi$ excitation is intrinsically broad; although the absolute uncertainty in the fitted peak position is sizable, it remains small compared to the peak linewidth.
In contrast, the higher-frequency $q=\pi/2$ mode becomes significantly sharper, thus, even a small absolute uncertainty in the fitted peak position corresponds to a larger fraction of the much narrower FWHM.

To probe many-body correlations beyond spectral features, we evaluate quantum Fisher information (nQFI), a well-established probe to lower bound the multipartite entanglement and remains accessible for large system sizes~\cite{hauke2016measuring, scheie2025tutorial} (See eq.~\ref{eq:nQFI}), and examine pairwise entanglement through the two-tangle $\tau_2$, extracted from short-range spin–spin correlations (Fig.~\ref{KCuF3_spectrum}I).
Both MPS and quantum simulations reveal at least four-partite entanglement based on nQFI in Fig.~\ref{KCuF3_spectrum}H, consistent with experimental observations.
The deviations at the $q=\pi$ point arise from its strong low-frequency weight and its sensitivity to long-time dynamics.
The two-tangle $\tau_2$ indicates a low degree of pairwise entanglement for KCuF$_3$, compared to the experimentally measured one-tangle $\tau_1$ of $0.76$~\cite{scheie2025tutorial}, which quantifies the single-site entanglement.

The low ratio $\tau_2/\tau_1$ indicates that only a small fraction of the total entanglement in KCuF$_{3}$ arises from pairwise correlations, consistent with a highly-entangled quantum-critical ground state. Notably, the $\tau_2$ evaluated based on the MPS simulations tends to overestimate $\tau_2$, which is also reported in Ref.~\cite{scheie2025tutorial}. While quantum noise suppresses both nQFI and $\tau_2$, the qualitative trends are preserved.

Together, these metrics demonstrate that quantum simulations capture not only the qualitative spectral features observed in INS, but also physically meaningful signatures of transport and entanglement. Entanglement measures extracted from the spectrum therefore serve as experimentally accessible bridge quantities linking real-space quantum correlations to universal many-body behavior.

\begin{figure*}
\centering
\includegraphics[width=0.98\textwidth]{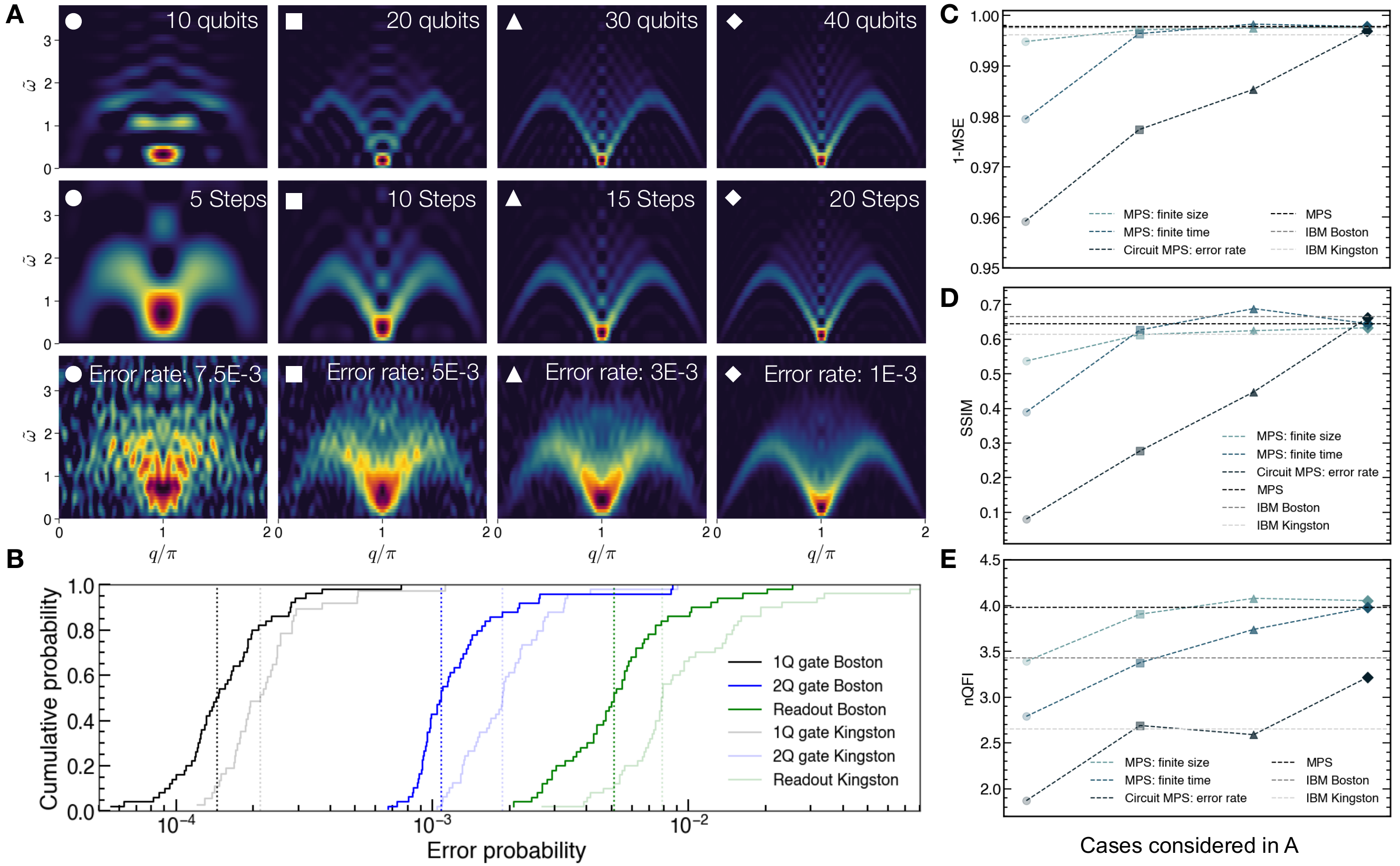}
\caption{ \label{error_analysis}
\textbf{Analysis of the spectrum with finite-size effects, finite-time effects, and depolarization errors.}
(\textbf{A}) MPS simulation of finite-size effects (first row), finite-time effects (second row), and noisy circuit-MPS simulation with depolarization errors (last row).
The baseline setup consists of 50 qubits, 20 time steps, and a fixed time step length of 0.6 and we vary either the system size or the number of time steps.
The noisy circuit-MPS results are simulated using a depolarization model applied only to two-qubit gates in the light-cone region of the quantum circuit.
(\textbf{B}) The cumulative distributions of gate errors for 50-qubit layout on $ibm\_kingston$ and $ibm\_boston$.
The median error rates of single-qubit gates are $2.130 \times 10^{-4}$ and $1.449 \times 10^{-4}$; those of two-qubit gates are $1.877 \times 10^{-3}$ and $1.080 \times 10^{-3}$; and those of readout are $7.876 \times 10^{-3}$ and $5.123\times 10^{-3}$.
Benchmarking MPS, noisy circuit-MPS, and quantum results for KCuF$_3$ using (\textbf{C}) MSE, (\textbf{D}) SSIM, and (\textbf{E}) nQFI. 
The markers and $x$-axes correspond to the cases considered in (\textbf{A}).
}
\end{figure*}

\subsection*{Hardware scaling of simulation accuracy}

To understand what enables current quantum devices to simulate INS spectra with sufficient resolution, we systematically analyze the roles of qubit number, circuit depth, and gate errors in determining spectral quality.
Using 50 qubits and 20 Trotter steps as a baseline, we vary each parameter independently and assess performance using representative metrics (MSE, SSIM, and nQFI) that capture spectral fidelity and many-body correlations.

The number of qubits and Trotter steps determine the momentum and frequency resolution, respectively. Small systems (10–20 qubits) yield heavily pixelated structure factors, particularly along the momentum axis (Fig.~\ref{error_analysis}A, first row), resulting in low SSIM and high MSE (Figs.~\ref{error_analysis}C,D) and making them poor proxies for real materials. 
In contrast, simulations with at least 40 qubits provide a clear momentum resolution, with MSE, SSIM, and nQFI approaching the 50-qubit MPS benchmark (Fig.~\ref{error_analysis}E). Similarly, finite evolution time produces pixelation along the frequency axis (Fig.~\ref{error_analysis}A, second row).
While MSE and SSIM largely converge after $\sim$10 Trotter steps, longer evolution with at least 20 steps is required to reliably capture the low-frequency correlations that dominate entanglement measures such as nQFI.

Access to large system sizes and long evolution times is ultimately limited by gate errors. 
To isolate the impact of noise, we model two-qubit gate errors within the light cone and evaluate the resulting spectra using MPS simulations. 
Reducing the effective error rate from $7.5\times10^{-3}$ to $1\times10^{-3}$ leads to marked improvements across all metrics, confirming that gate fidelity is the primary bottleneck.

This trend is directly reflected in hardware experiments on two generations of IBM Heron processors, \textit{ibm\_kingston} and \textit{ibm\_boston} (Fig.~\ref{error_analysis}B).  The lower two-qubit error rates achieved on \textit{ibm\_boston} reach the threshold required for the quantitative and qualitative agreement with INS spectra demonstrated in this work. As median two-qubit error rates approach $10^{-3}$, our results indicate that current digital quantum processors have entered a regime in which quantitatively meaningful INS simulations become feasible. 
Continued advances in error suppression and mitigation are therefore expected to extend this capability to increasingly complex spectra. Furthermore, the inherent flexibility of Hamiltonian choice on universal digital platforms enables the exploration of non-integrable dynamics that remain challenging for both classical methods and equilibrium-centric analog simulations~\cite{leclerc2026one}.

\begin{figure}
\centering
\includegraphics[width=1\linewidth]{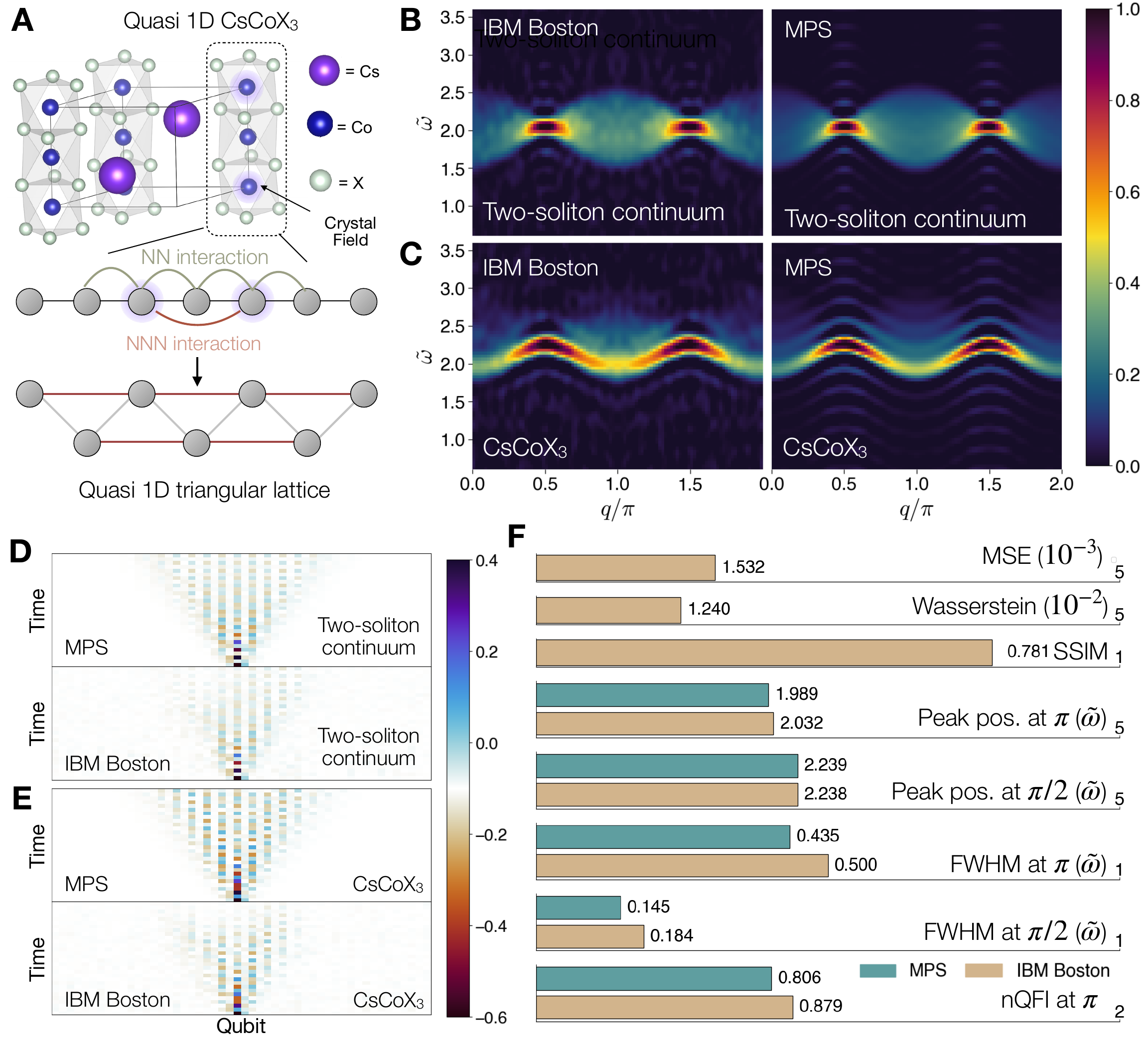}
\caption{
\textbf{INS spectrum for two-soliton continuum and CsCoX$_3$ model. }
(\textbf{A}) Crystal structure of CsCoX$_3$, which requires NNN interactions with 1-D triangular geometry for realistic modeling.
Comparison of the INS spectrum and RGF for 50-qubit MPS and quantum simulations with (\textbf{B,D}) only nearest-neighbor (NN) interactions and (\textbf{C,E}) additional next-nearest-neighbor (NNN) interactions.
Without NNN interactions, the spectrum exhibits a bow-tie structure consistent with the theoretical two-soliton continuum. 
Upon including NNN interactions, bow-tie shape evolves into a lower-band dispersion.
We used a time-step length $\Delta t = 0.8$, and 30 first-order trotter steps and the quantum runtimes for both cases were $\sim 18$ minutes.
(\textbf{F}) Benchmarking the quantum spectrum against MPS results using various metrics.}
\label{CsCoX3_spectrum}
\end{figure}

\subsection*{Beyond Integrable Hamiltonians}
Leveraging this capability, we move beyond the analytically solvable limit of KCuF$_3$ to a non-integrable extension of the XXZ model featuring ferromagnetic NNN interactions. These longer-range couplings effectively extend the strictly 1D chain toward a quasi-1D triangular geometry and capture essential features of interchain interactions in the CsCo$X_3$ compounds above their N\'eel temperature~\cite{matsubara1991magnetic,matsubara1991two,shiba2003exchange}, as illustrated in Fig.~\ref{CsCoX3_spectrum}A. The Hamiltonian is
\begin{align}
\label{NNN}
    H_{\text{NNN}} &= H_{\text{NN}}- 2J'\sum_{i=1}^{N-2} \left[
    S_i^Z S_{i+2}^Z +
    \epsilon'(S_i^X S_{i+2}^X + S_i^Y S_{i+2}^Y)
    \right]
\end{align}
where $J'$ and $\epsilon'$ denote the exchange interaction strength and anisotropy associated with the NNN couplings. 
We set $J'/J \approx 0.095$ and $\epsilon=\epsilon'=0.145$, which has been reported to accurately model CsCoCl$_3$ above its N\'eel temperature~\cite{matsubara1991magnetic}. While alternative descriptions invoke unequal anisotropies for NN and NNN couplings~\cite{shiba2003exchange} or staggered fields~\cite{nagler1983ising}, these extensions can be straightforwardly incorporated into our circuit construction without increasing the two-qubit gate depth. We therefore focus on this minimal parameter choice to isolate the role of longer-range interactions.

To isolate the effects of NNN couplings, we first examine the spectrum in the absence of NNN coupling. 
This lies close to the pure Ising limit and remains analytically tractable, allowing direct comparison with the non‑integrable case. With strong ZZ anisotropy, the DSF exhibits a broad, nearly uniform distribution between the upper and lower edges of a gapped bow-tie structure, a hallmark of diffusive transport with $t^{-1/2}$ correlation decay and consistent with the well-known two-soliton continuum~\cite{goff1995exchange} (Fig.~\ref{CsCoX3_spectrum}B). 
Introducing the transverse coupling $\epsilon(S^X S^X+S^Y S^Y)$ lifts the degeneracy of the Ising ground states and generates continuum of excited states around $2J$~\cite{ goff1995exchange}. 
Quantum simulations faithfully capture this theoretical continuum (Fig.~\ref{CsCoX3_spectrum}B).
The corresponding light-cone (Fig.~\ref{CsCoX3_spectrum}D) is comparatively narrower than in KCuF$_3$, reflecting slower quasiparticle propagation in the diffusive regime.

Including ferromagnetic next-nearest-neighbor (NNN) interactions qualitatively reshapes the two-soliton continuum into a single magnon mode, as shown in Fig.~\ref{CsCoX3_spectrum}C.
The lower edge of the spectrum is enhanced, reflecting a ``bound" dispersion favored by NNN coupling and the suppression of higher-energy ``anti-bound" dispersion.
Crucially, the weak NNN couplings introduce controlled departures from integrability and break analytical tractability. 
The lack of analytical benchmark makes direct comparisons between MPS and quantum simulations particularly meaningful. 
We find good overall agreement: peak positions from quantum simulations at $q=\pi/2$ and $q=\pi$ lie within a 10\% error relative to the FWHM of the MPS (Fig.~\ref{CsCoX3_spectrum}F), and the nQFI remains below 1 at $q=\pi$, indicating a weakly entangled regime with low multipartite entanglement consistent with theoretical expectations (Fig.~\ref{CsCoX3_spectrum}F). 
Together, these results demonstrate that quantum simulations can reliably capture the qualitative spectral features of non-integrable quasi-one-dimensional magnets, motivating future experimental verification via INS.

\subsection*{Conclusion and Outlook}

By reconstructing the spectral signatures of KCuF$_3$ and non-integrable models of CsCoX$_3$, we demonstrate a framework widely applicable across a broad class of quasi-1D quantum magnets. This methodology covers diverse exchange ($J$) and anisotropy ($\epsilon$) parameters represented by materials such as Cs$_2$CoCl$_4$~\cite{cs2cocl4} and BaCo$_2$V$_2$O$_8$~\cite{baco2v2o8}. The quantitative accuracy demonstrated in this work stems from the direct mapping between INS and quantum simulation workflows, supported by improved gate fidelities on pre-fault-tolerant hardware. 
This achievement represents a substantial advance over prior quantum calculations of DSFs. Notably, the suite of metrics introduced in this work provides a rigorous framework for tracking the evolution of quantum simulation into an indispensable tool for condensed-matter physics. These metrics are readily adaptable to other experimental probes, such as Raman scattering. As hardware scales to address classically intractable problems, such as spin dynamics in the 2D triangular antiferromagnet, this framework enables quantum processors to compare directly with lab experiments~\cite{scheie2026nonlinearlightconespreading}. 
By utilizing real materials as a primary benchmark without a classical intermediary, this approach establishes a definitive path for the use of quantum computers as practical tools for scientific discovery.


\section*{Acknowledgments}
B.P. would like to thank Abhinav Deshpande and James Raftery for useful discussions. This research was supported by the U.S. Department of Energy, Office of Science, National Quantum Information Science Research Centers, Quantum Science Center. 
The work at the University of Illinois at Urbana-Champaign (UIUC) is supported by the Taiwan UIUC Scholarship under the official memo No.\ 1100063269M and by the IBM Illinois Discovery Accelerator Institute (IIDAI).
A portion of this research used resources at the Spallation Neutron Source, a DOE Office of Science User Facility operated by the Oak Ridge National Laboratory under IPTS-34938 and IPTS-26614. 
This work made use of the Illinois Campus Cluster, a computing resource that is operated by the Illinois Campus Cluster Program (ICCP) in conjunction with the National Center for Supercomputing Applications (NCSA) and which is supported by funds from UIUC, as well as the Gilberth GPU clusters of Rosen Center for Advanced Computing (RCAC) Purdue University. 

\bibliographystyle{apsrev}
\bibliography{main}

\beginsupplement
\newpage
\begin{center}
\textbf{\large Benchmarking quantum simulation with neutron-scattering experiments \\\vspace{0.3 cm}}

Yi-Ting Lee $^{1\ast}$, Keerthi Kumaran $^{2,3\ast}$, Bibek Pokharel$^{3,4\dagger}$, Allen Scheie$^{5}$, Colin L. Sarkis$^{6}$, Stephen E. Nagler$^{6}$, D. Alan Tennant$^{7,8}$, Travis S. Humble$^{3}$, Andr\'e Schleife$^{1,9}$, Abhinav~Kandala$^{3,4}$, and Arnab~Banerjee$^{2,3\dagger}$

$^1$\textit{Department of Materials Science and Engineering, University of Illinois at Urbana-Champaign, Urbana, Illinois 61801, USA}

$^2$\textit{Department of Physics and Astronomy, Purdue University, West Lafayette, Indiana 47907, USA}

$^3$\textit{Quantum Science Center, Oak Ridge National Laboratory, Oak Ridge, Tennessee 37831, USA}

$^4$\textit{IBM Quantum, IBM Thomas J. Watson Research Center, Yorktown Heights, New York, USA}

$^5$\textit{MPA-Q, Los Alamos National Laboratory, Los Alamos, New Mexico 87545, USA}

$^6$\textit{Neutron Scattering Division, Oak Ridge National Laboratory, Oak Ridge, Tennessee 37831, USA}

$^7$\textit{Department of Physics and Astronomy, University of Tennessee, Knoxville, Tennessee 37996, USA}

$^8$\textit{Department of Materials Science and Engineering, University of Tennessee, Knoxville, Tennessee 37996, USA}

$^9$\textit{Materials Research Laboratory, University of Illinois at Urbana-Champaign, Urbana, Illinois 61801, USA}

\end{center}

\begin{figure*}[b]
\centering
\includegraphics[width=1\textwidth]{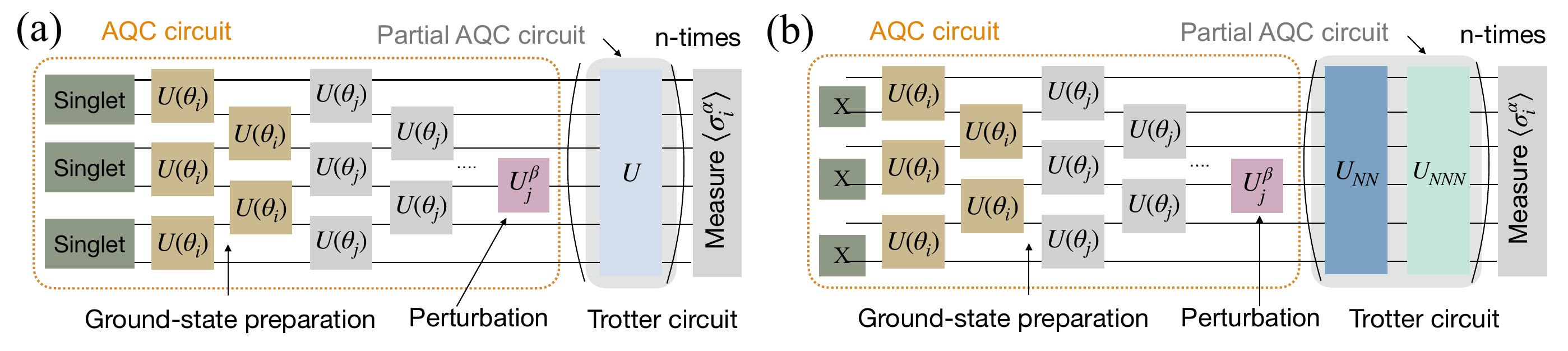}
\caption{\label{circuit} 
\textbf{ Quantum circuit for measuring the dynamical structural factor}.
\textbf{A} Quantum circuit for XX model and KCuF$_3$, where $U$ refers to second-order trotterization.
\textbf{B} Quantum circuit for Ising-like XXZ model with NNN interactions, where the initial state is the Néel state instead of the singlet product state. } 
\end{figure*}

\subsection{Spin Hamiltonian}
For KCuF$_3$, the ratio of interchain to intrachain interaction is only 0.027~\cite{ikeda1973neutron}.
At the N\'eel temperature of 39~K, the system is dominated by intrachain behavior and can be effectively modeled by the 1D Heisenberg model (equivalent to the XXZ Model at the isotropic point, \(\epsilon = 1\))~\cite{lake2013multispinon}.
Although the low-energy spectrum (\(<30\)~meV) is expected to be influenced by interchain exchange below the N\'eel temperature, the experimental spectrum at 6~K remains consistent with the Bethe ansatz prediction at 0~K \cite{lake2013multispinon}. 
Therefore, in this work, we simulate the DSF of the 1D XXZ model at the isotropic point, as shown in Figure~\ref{overview}, and benchmark it against the corresponding neutron scattering spectra reported in Ref.~\cite{lake2013multispinon} with the quantum circuit in figure~\ref{circuit}(a).

\subsection{Retarded Green's function}
The retarded Green's function (RGF) takes the following form in coordinate space (space and time).
\begin{equation}
\label{eq:Green}
G^R_{\alpha, \beta}(i,j,t) = -\frac{i}{2}\langle S^\alpha_i(t) S^\beta_j - S^\beta_j S^\alpha_i(t)\rangle_0
\end{equation}
where $S^{\alpha}_i$ are the spin-$\frac{1}{2}$ operators acting on $i^{\text{th}}$ spin and $\alpha, \beta \in X, Y, Z$ are the Pauli spin operators.
When the system is close to the thermodynamic limit, one can approximate the computation in Eq.\ \ref{eq:Green} by fixing \( j \) to the central site \( j_c \) \cite{bauer2025progress}.
This reduces the number of required circuits by a factor equal to the system size, since there is no need to iterate \( j \) over all qubit sites.

\subsection{INS experiment}

The KCuF$_3$ spectrum at T~=~6~K was measured and reported in Ref.~\cite{lake2013multispinon} using the thermal neutron time-of-flight MAPS thermal neutron spectrometer at the ISIS Facility at, Rutherford Appleton Laboratory, UK.
The KCuF$_3$ spectrum at T~=~75~K was measured and reported in Refs. \cite{scheie2022quantum} using a combination of MAPS and SEQUOIA time-of-flight thermal neutron chopper spectrometer at Oak Ridge National Laboratory (ORNL): MAPS for high-energy ($\hbar \omega > 6$~meV), SEQUOIA for low-energy ($\hbar \omega < 6$~meV). The $c$-axis was oriented in the scattering plane perpendicular to the incident beam. Data were corrected for the Cu$^{2+}$ form factor and integrated over all directions perpendicular to the chains. Because KCuF$_3$ has weak magnetic order below its ordering temperature $T_N = 39$~K, there are subtle differences between the longitudinal and transverse responses at low energy $\tilde{\omega} < 0.5$ \cite{PhysRevB.71.134412}, but these are minor effects and the spectrum overall matches very closely to the perfect 1D isotropic model  \cite{lake2013multispinon}.

\subsection{Implementation details for quantum computing}
\label{quantum_implementation_details}
To measure the DSF, we follow the procedure shown in Figure~\ref{overview}, beginning with variational ground-state preparation~\cite{yu2023simulating}.
For KCuF$_3$, the state preparation begins with a product of singlet states concatenated with an ansatz that incorporates a time-evolution circuit structure, as shown in Figure~\ref{circuit}(a).
Since the ground state of KCuF$_3$ belongs to the class of quantum critical states, the construction of quantum circuits that accurately capture its highly entangled nature is computationally demanding.
Therefore, we target a fidelity of 80\% for the ground-state circuit preparation.
For the Hamiltonian in the Ising limit with strong ZZ anisotropy,
the ansatz to prepare the ground state starts from the N\'eel state rather than the singlet product state as shown in Figure~\ref{circuit}(b).
Because the ground state is expected to be less entangled than that of KCuF$_3$, we target fidelity exceeding 90\%.

To extract the RGF $G^R_{\alpha, \beta}(q, \omega) $, 
we implement the protocol in Ref.~\cite{baez2020dynamical} and utilize the center-site approximation~\cite{bauer2025progress} to reduce the cost of measurements as shown in Figure~\ref{overview}. 
Specifically, we first apply a local rotation gate at the center site ($j_c$):
\begin{equation}
\label{perturb}
    U_{j_c} = \frac{1}{\sqrt{2}} (1 - i \sigma_{j_c}^\beta)
\end{equation}
After time evolution $U(t)= e^{-iHt}$, 
the state becomes $|{\psi(t)} \rangle = U(t) U_{j_c} |{\psi_{GS}} \rangle$, and measuring $\sigma_i^\alpha$ yields:
\begin{equation}
    G^{R}_{\alpha, \beta}(j,j_c,t) = \langle \psi(t)|\sigma^\alpha_j|\psi(t)\rangle
\end{equation}
This expression is simplified by the parity symmetry of the 1D XXZ model. 
The RGF $G^{R}_{\alpha,\beta}(q,\omega)$ is obtained by taking the Fourier transform of $G^{R}_{\alpha,\beta}(j, j_c, t)$ as follows:
\begin{equation}
G^R_{\alpha,\beta}(q,\omega) = \sum^n_{j=0} \sum^N_{k=0} e^{-iq(j-j_c)} e^{i \omega k \Delta t} G^R_{\alpha,\beta}(j,j_c,t) 
\end{equation}
where $n$ is the number of spins, $N$ is the number of time steps, and $\Delta t$ is the time interval for trotterization.
The center-site approximation reduces the number of circuits measured by a factor of $n$, since we only perturb the central site once, compared to perturbing each qubit individually with $n$ separate circuits.
The DSF is the Fourier transform of the RGF.
Since we use 50 qubits, an even number, the RGF preserves the mirror symmetry of the ground state under the center-site approximation, i.e., mirror symmetry upon perturbing the $n//2-1$ and $n//2$ sites.
We therefore average the final spectrum with its inverted counterpart.
We also note that the RGF can also be evaluated from the imaginary part of the spin-spin autocorrelation function, resulting in an identical circuit construction following the strategy in Ref.~\cite{mitarai2019methodology, lee2026digital}.

To implement the time-evolution operator on quantum computers, we employ second-order Trotterization for KCuF$_3$ and the XX model.
Because implementing the time-evolution operator with NNN  (NNN) interactions remains costly on near-term devices due to the limited connectivity of current qubit topologies, we apply first-order Trotterization, which involves the implementation of swap layers.

To reduce the circuit depth for long-time simulations on current devices, we employ approximate quantum compilation (AQC) based on the tensor network approach~\cite{robertson2025approximate}.
We optimized the quantum circuit over a short time period and achieved circuit fidelity of 90 \% compared to the time-evolved state obtained from matrix-product-state (MPS) simulation using a bond dimension of 128.
For longer-time simulations that cannot be accurately approximated by just AQC, we concatenate Trotter circuits to the AQC circuit, keeping the ratio of two-qubit gates in the AQC part below 50 \% of the deepest simulated circuit.

For the quantum experiment, we perform our simulation on the IBM Heron processor $ibm\_boston$, which comprises 156 fixed-frequency transmon qubits~\cite{koch2007charge}, featuring heavy-hex connectivity.
We perform dynamical decoupling (DD) using the XY4 sequence~\cite{maudsley1986modified} and employ 1000 Pauli-twirled (PT) circuits~\cite{wallman2016noise} with 128 shots each to suppress coherent noise.
We further apply twirled readout error extinction (TREX)~\cite{van2022model} to mitigate measurement errors.
Circuit compilation, the implementation of error mitigation and error suppression are handled by \textsc{Qiskit}~\cite{javadiabhari2024quantumcomputingqiskit}.

\begin{figure*}[b]
\centering
\includegraphics[width=1\textwidth]{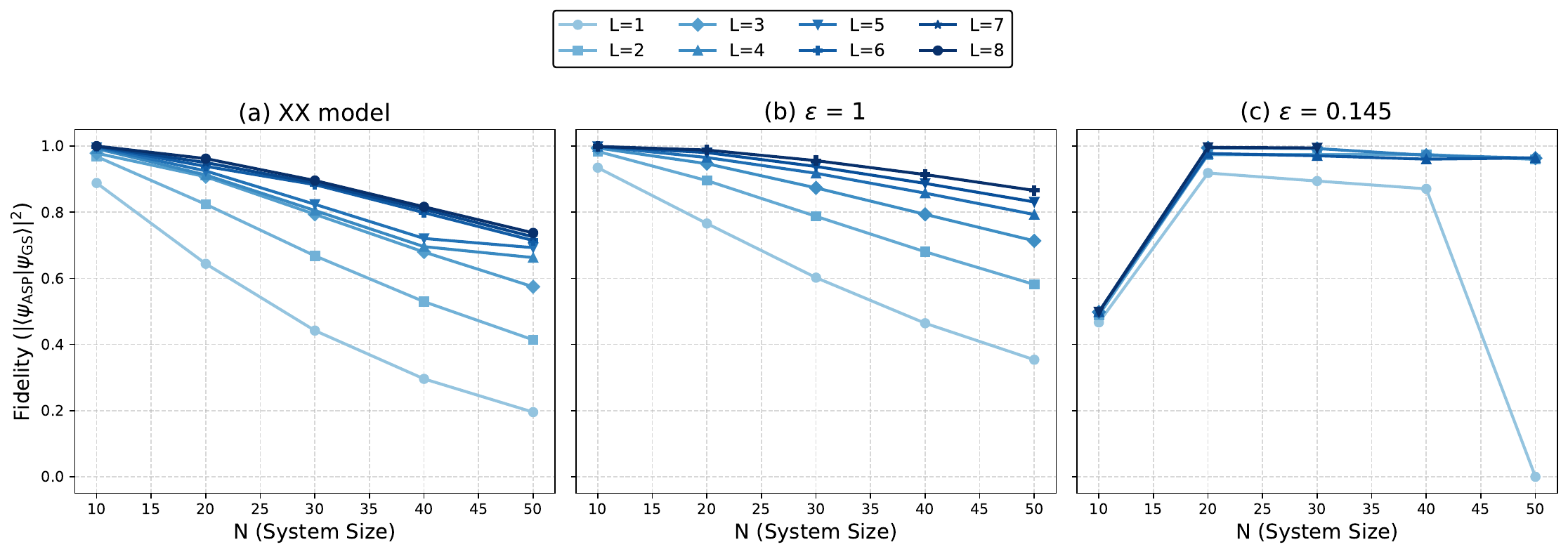}
\caption{\textbf{Fidelity of variational state preparation for the XXZ spin models considered in this work as a function of system size $N$ and the number of ansatz layers $L$.} Panel \textbf{A} shows results for the XX model, panel \textbf{C} for $\epsilon = 1$ and \textbf{B} for $\epsilon = 0.18$. We observe that between the XX model and the $\epsilon = 1$, the ansatz achieves higher fidelities with fewer layers in the latter. In both these cases, the achieved fidelities for a given number of ansatz layers decreases with increasing system size. The trends are noticeably different for \textbf{C} $\epsilon = 0.145$, characterized by the rapid saturation of fidelities near $1.0$.
This behavior may be attributed to our different choice of initial state in \textbf{C}; notably, the initial N\'{e}el state has a significantly higher overlap with the true ground state compared to the initial states used in cases \textbf{A} and \textbf{B}.}
\label{fig:state_prep}
\end{figure*}

\subsection{Ground state preparation} 
\label{supp:ASP}
To prepare the ground-state circuits for the spin models studied in Figures~\ref{KCuF3_spectrum} and~\ref{CsCoX3_spectrum}, we perform optimization of Hamiltonian variational ansatz acting on appropriate initial states. The optimization is performed classically, with the loss function chosen to be the fidelity overlap with respect to the MPS ground state computed using DMRG, i.e., $|\langle \psi_o | \psi_{\mathrm{GS}} \rangle|^2$.
For the 1D XX model and the 1D XXZ model at the isotropic point, we adopt the procedure outlined in the work \cite{yu2023simulating}. The process begins with pairs of singlets, which form the ground state of a Hamiltonian corresponding to the XXZ model acting only on the odd pairs of the system, 
then we apply the variational Hamiltonian ansatz on the entire system. 
This approach enables convergence to the ground state for both cases. 
For the ground states corresponding to $\epsilon = 0.145 $, the N\'eel state is closer to them than the singlet configuration.
Therefore, it is more appropriate to start from the N\'eel state and use a variational Hamiltonian ansatz.
Using this approach, we achieve significantly higher fidelities compared to starting from singlet states, reaching approximately $95.9\%$.

To evaluate the quality of the ansatz used in the ground-state preparation \cite{yu2023simulating}, we plot the converged fidelity, $|\langle \psi_o | \psi_{\mathrm{GS}} \rangle|^2$, against the system size $n$ for different numbers of ansatz layers, ranging from $L = 1$ to $L = 6$ (see Figure~\ref{fig:state_prep}). The parameter initialization in our variational ground-state preparation for a given number of ansatz layers proceeded as follows: compute converged parameters for system size $n = 10$, use those as the starting point for $n = 20$, then use the converged parameters corresponding to $n = 20$ as the starting point for $n = 30$, and so on, until $n = 50$.

Although the fidelity decreases with system size, this behavior is expected because both the XX and XXZ models lie in critical phases in the thermodynamic limit. However, for finite systems up to 50 qubits, as considered in this work, even with just six ansatz layers, we obtain fidelities of approximately $70\%-75\%$ for the XX model and close to $85\%$ at the isotropic point of the XXZ model. These fidelities are sufficient to accurately capture the growth of the retarded Green's function over the simulated time window and to reproduce the corresponding DSF spectra. This further highlights that reliable DSF simulation does not require systems much larger than $\sim 100$ qubits.

For the simulation results presented in Figure~\ref{KCuF3_spectrum}, we used five layers.
We note that for the simulations shown in Figure~\ref{CsCoX3_spectrum}B, we used the same ground state as in the case without NNN interactions. 
This choice is justified because these NNN interactions are very small in magnitude, and the fidelity overlap with respect to the true ground state (including NNN interactions) is nearly identical to that obtained without them.
The small NNN interaction does not significantly alter the ground-state wavefunction, as the fidelity between the states with and without the NNN interaction is $99.4 \%$.

\subsection{Time evolution circuit}
\begin{figure*}
\centering
\includegraphics[width=0.9\textwidth]{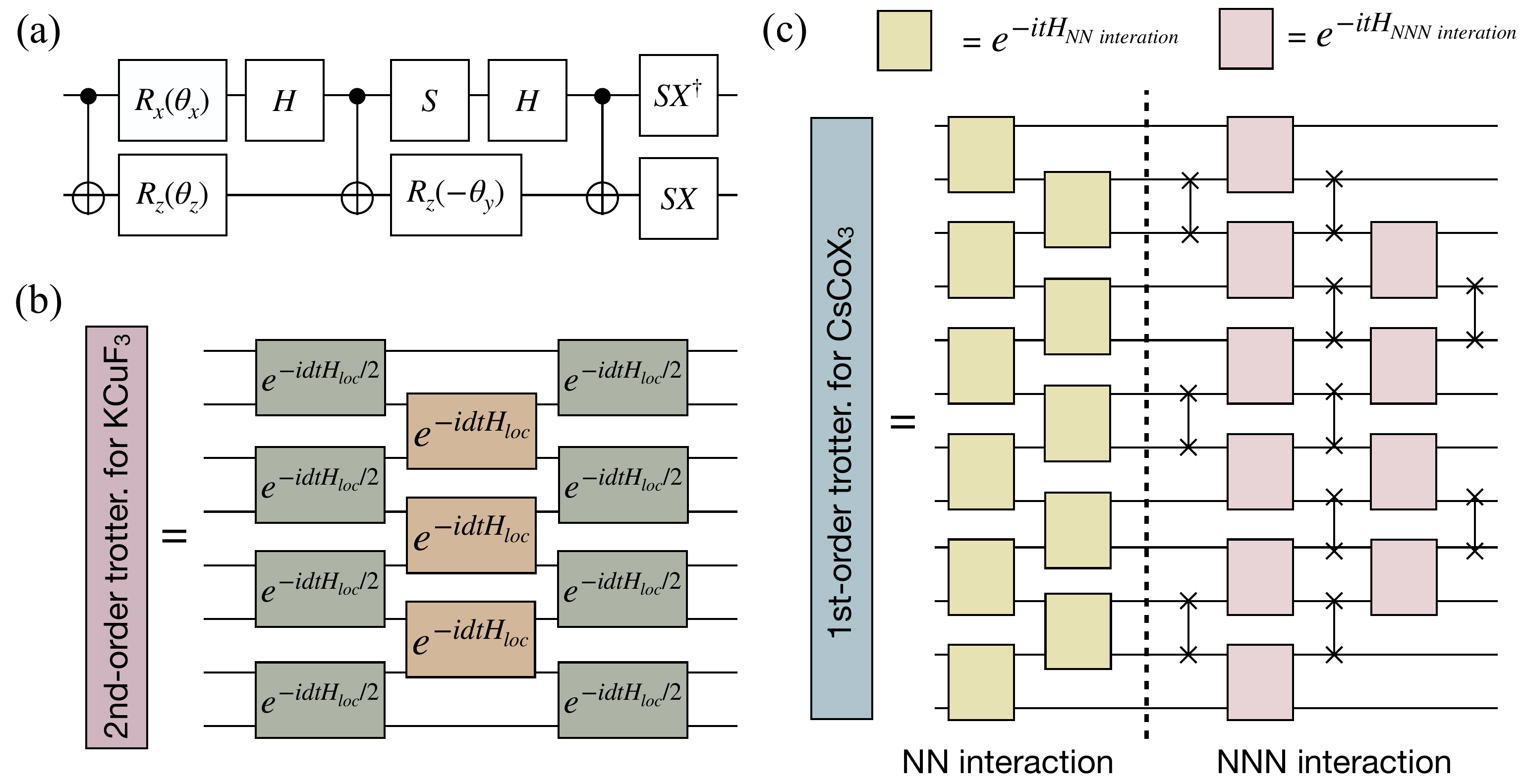}
\caption{\label{trotter_circuit} 
\textbf{Circuit implementation for the trotterization. }
(a) Quantum circuit for the local time evolution operator with $S^X_i S^X_{i+1} $, $S^Y_i S^Y_{i+1}$, and $S^Z_i S^Z_{i+1}$
interactions, where the parameters $\theta_x, \theta_y, \theta_z$ are adjusted according to the strengths of the individual interactions and the time step length. 
(b) The circuit implementation of the second-order trotterization for the XX model and KCuF$_3$, where $H_{loc}$ denotes the local Hamiltonian $J(S^X_j S^X_{j+1} + S^Y_j S^Y_{j+1} + S^Z_j S^Z_{j+1})$ at the isotropic point.
(c) The circuit implementation of the first-order trotterization for Ising-like XXZ model with NNN interactions (Eq.~\ref{NNN}).
To implement the NNN interaction within the 1D model on IBM quantum computers, we employ SWAP gates to bring next-nearest neighbor pairs adjacent to each other and perform the gate operation shown in (a) using the parameters corresponding to the NNN interaction.
} 
\end{figure*}

In this study, we employ trotterization to implement the time-evolution operator in quantum circuits.
For the time-evolution circuit of KCuF$_3$, we employ second-order Trotterization based on the Suzuki-Trotter formulas \cite{hatano2005finding}.
\begin{equation}
\label{2nd_trotter}
 e^{-iHt} \approx  \left[\prod_{j=1}^N e^{-i \frac{t}{2n}h_j} \prod_{j=N}^{1} e^{-i \frac{t}{2n}h_j}\right]^n 
\end{equation}
where $n$ and $t$ refer to the number of Trotter steps and the total time in the simulation, respectively.
For KCuF$_3$, $h_j$ denotes the minimal set of non-commuting operators with nearest-neighbor interaction, which can generally be expressed in terms of the linear combination of  $S^X_j S^X_{j+1}$, $S^Y_j S^Y_{j+1}$, and $S^Z_j S^Z_{j+1}$ pauli operators and the circuit implementation is shown in Figure~\ref{trotter_circuit}(a).
The second-order Trotterization layout for KCuF$_3$ is illustrated in Figure~\ref{trotter_circuit}(b).

For the Hamiltonian in Equation \ref{NNN}, since it involves NNN interactions, performing second-order trotterization on the current device becomes computationally unfeasible. 
Therefore, we employ the first-order trotterization via the lie-trotter formula \cite{lloyd1996universal}.
\begin{equation}
\label{1st_trotter}
 e^{-iHt} \approx  \left[\prod_{j=1}^N e^{-i \frac{t}{n}h_j} \right]^n  = \left[\prod_{j=1}^N e^{-i \frac{t}{n} H_{NN}} e^{-i \frac{t}{n} H_{NNN}} \right]^n
\end{equation}
where it can be further represented as the product of the time-evolution operators for the nearest-neighbor (NN) term $H_{\mathrm{NN}}$ and the NNN term $H_{\mathrm{NNN}}$, corresponding to the first and second terms in Equation~\ref{NNN}, respectively.
The implementation of the first-order Trotterization is shown in figure~\ref{trotter_circuit}(c), where the circuit is composed of time-evolution operators based on NN and NNN interactions.
For the time evolution under $H_{NNN}$, SWAP gates are employed to implement two-qubit interactions across three qubits, addressing the limitations of the current qubit topology, which is also reported in Ref~\cite{chowdhury2024enhancing}.

\subsection{Approximate quantum compiling}
We employ AQC based on a tensor network approach~\cite{robertson2025approximate} to reduce two-qubit gate depth and enable simulations for longer times. Specifically, for short-time evolution, the quantum circuit can be simulated efficiently on classical hardware because the entanglement can still be accurately described using MPS simulation.
The original quantum circuit is approximated by optimizing the ansatz parameters to achieve the target fidelity, using a shallower circuit as the ansatz.
However, as entanglement grows with simulation time, the required number of quantum gates increases, and the cost of MPS simulations becomes significantly more computationally expensive due to the memory demands associated with optimizing highly entangled quantum states.
Thus, simulation at long-time period need to be achieved by concatenating the AQC circuits with the Trotter circuits.

To prepare the AQC circuit, we initiate the process using the original circuit with fewer Trotter steps and employ KAK decomposition to generate the ansatz~\cite{khaneja2000cartan}, the target fidelity is set to 90 \%. 
Thus, the difference between the ansatz and the target circuit is the number of Trotter steps. This discrepancy is expected to decrease over time, as the number of Trotter steps required to accurately approximate an entangled quantum state typically increases with simulation time.
The AQC process terminates once the optimization process requires more than 500 GB of memory, which typically exceeds the memory of a single computational node.
The time-evolution quantum circuit beyond the AQC capability will be described by a Trotter circuit, as mentioned beforehand.
The generation of the ansatz and the optimization process are carried out by~\textsc{qiskit$\_$addon$\_$aqc$\_$tensor}~\cite{qiskit-addon-aqc-tensor} package.

\begin{figure*}
\centering
\includegraphics[width=0.9\textwidth]{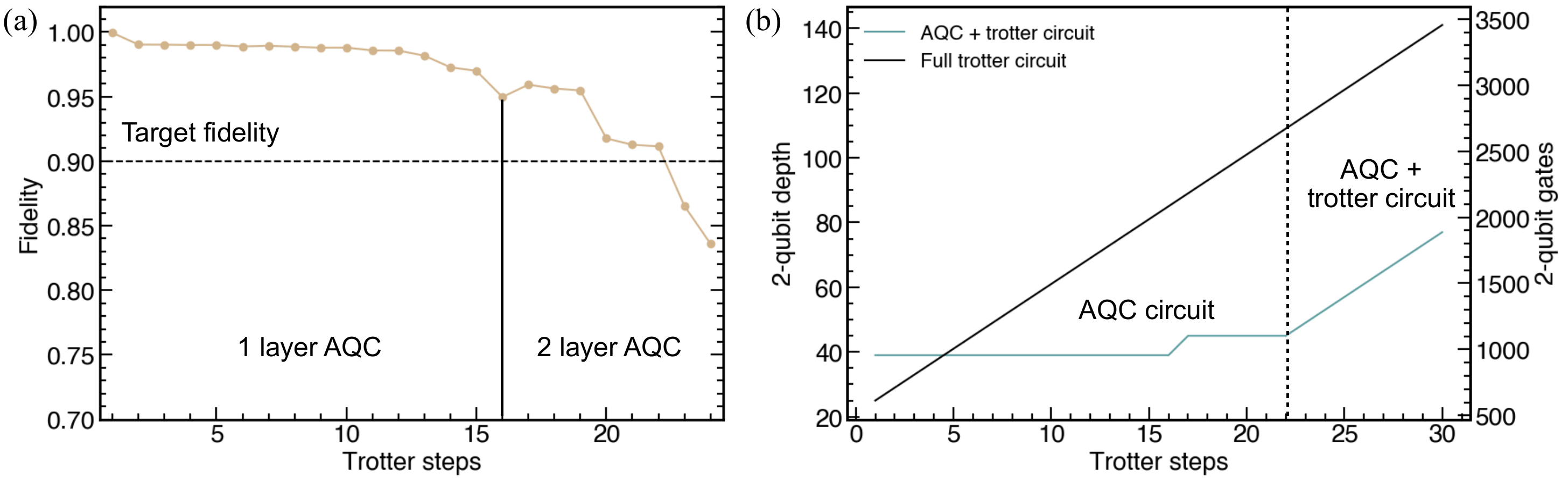}
\caption{\label{aqc_xx}
\textbf{Approximate Quantum Compiling (AQC) details for the XX model simulations.}
(a) Fidelity between the original Trotter circuit and the AQC circuit.
The deepest ansatz we employ consists of five Trotter layers combined with the ground-state circuit.
(b) Comparison of the 2-qubit depth and gate count between the original Trotter circuit and the AQC+Trotter circuit. 
After 22 Trotter steps, the depth and gate costs increase linearly as the Trotter circuit is concatenated to perform the simulation over 22 steps.} 
\end{figure*}

For the XX model, we start with the 5-layer ground state with 760 two-qubit gates and depth of 31. 
The fidelity of the AQC circuit with different number of trotter steps is shown in figure~\ref{aqc_xx}(a).
Based on the target fidelity, we perform AQC circuit to simulate the system up to 10 Trotter steps. 
The deepest ansatz we can employ corresponds to the original circuit consisting of two Trotter steps, for which the AQC optimization of the ansatz with 3 Trotter steps requires over 700 GB. 
A comparison of the 2-qubit gate count and circuit depth between the circuit with and without AQC are shown in figure~\ref{aqc_xx}(b). 
The deepest circuit after AQC reaches a 2-qubit depth of 77 and contains 1888 two-qubit gates.
Compared to the original circuit, the number of two-qubit gates is reduced by 1567 and the two-qubit depth is reduced by 64 .
The two-qubit gates from the AQC circuit account for 50.7 \% of those in the deepest quantum circuit.
Notably, the 2-qubit depth and gate count are higher for the AQC circuit than for the full Trotter circuits before 5 trotter steps.
This is a consequence of the~\textsc{qiskit$\_$addon$\_$aqc$\_$tensor}~\cite{qiskit-addon-aqc-tensor} package employing KAK decomposition when generating the ansatz circuit for the time evolution block of the XX model. Since the XX model's time-evolution block contains only two CNOT gates, preparing the ansatz with XX, YY, and ZZ interactions in the package requires three CNOT gates, leading to an increased circuit depth.
Besides, the use of a 6-layer ansatz required at least 700 GB of memory just to initiate the AQC process, validating the choice of the 5-layer ansatz for subsequent runs.

\begin{figure*}
\centering
\includegraphics[width=0.9\textwidth]{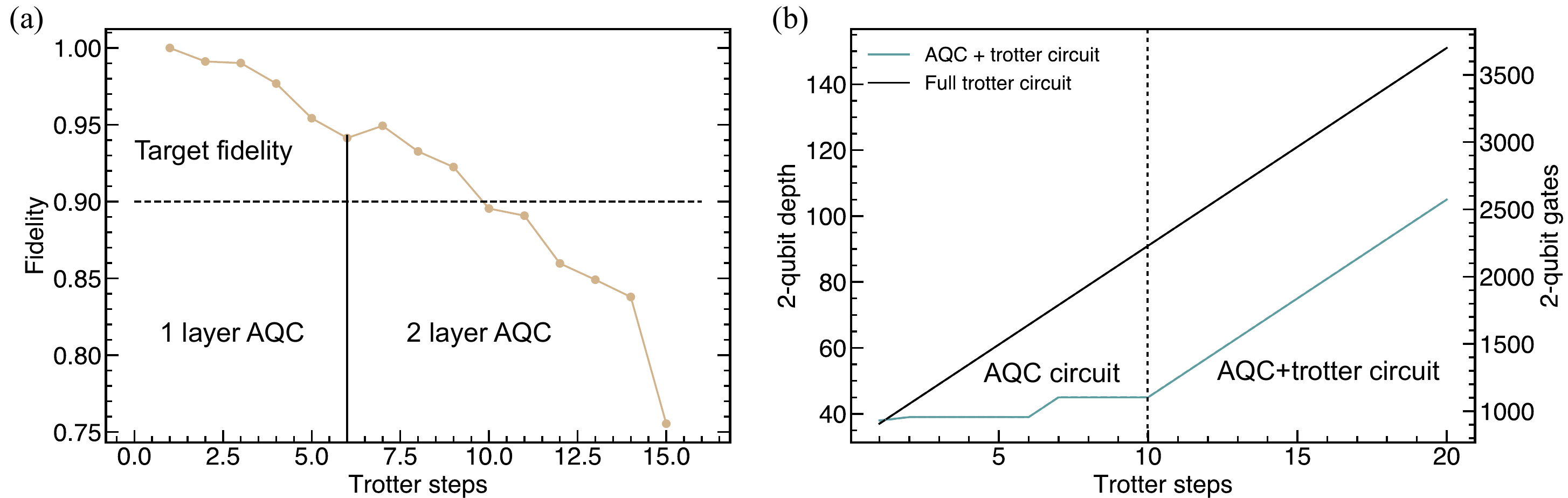}
\caption{\label{aqc_kcuf3}
 \textbf{Approximate Quantum Compiling (AQC) details for the KCuF$_3$ simulations.}
(a) Fidelity between the original Trotter circuit and the AQC circuit.
Due to memory limitations, the deepest ansatz we employ consists of two Trotter layers combined with the ground-state circuit.
(b) Comparison of the 2-qubit depth and gate count between the original Trotter circuit and the AQC+Trotter circuit. 
After 10 Trotter steps, the depth and gate costs increase linearly as the Trotter circuit is concatenated to perform the simulation over 10 steps.} 
\end{figure*}

For KCuF$_3$, we start with the entangled ground-state circuit, which has 760 two-qubit gates and depth of 31, and gradually increase the Trotter steps in the ansatz to approximate the full quantum circuit. 
The fidelity of the AQC circuit with different number of trotter steps is shown in figure~\ref{aqc_kcuf3}(a).
Based on the target fidelity, we select AQC circuit for original circuits with up to 10 Trotter steps, achieving a fidelity above 89.5 \%. 
Although it does not reach the 90 \% criteria, we find it beneficial to sacrifice a small amount of fidelity in order to reduce the circuit depth and the number of two-qubit gates.
Notably, the deepest ansatz we can employ corresponds to ground state circuit with 2 Trotter steps, for which the optimization of the entangled state over 3 Trotter steps requires over 850 GB of memory for the case of 10 Trotter-steps evolution. 
To simulate time evolution beyond 10 Trotter steps, we concatenate the standard Trotter circuit with the AQC circuit.
A comparison of the 2-qubit gate count and circuit depth between the circuit with and without AQC is shown in figure~\ref{aqc_kcuf3}(b). 
The deepest circuit after AQC reaches a 2-qubit depth of 103 and contains 2,574 two-qubit gates.
Compared to the original circuit, the number of two-qubit gates is reduced by 1,126, and the two-qubit depth is reduced by 46.
The two-qubit gates in the AQC circuit account for 42.6 \% of those in the deepest quantum circuit.

\begin{figure*}
\centering
\includegraphics[width=0.9\textwidth]{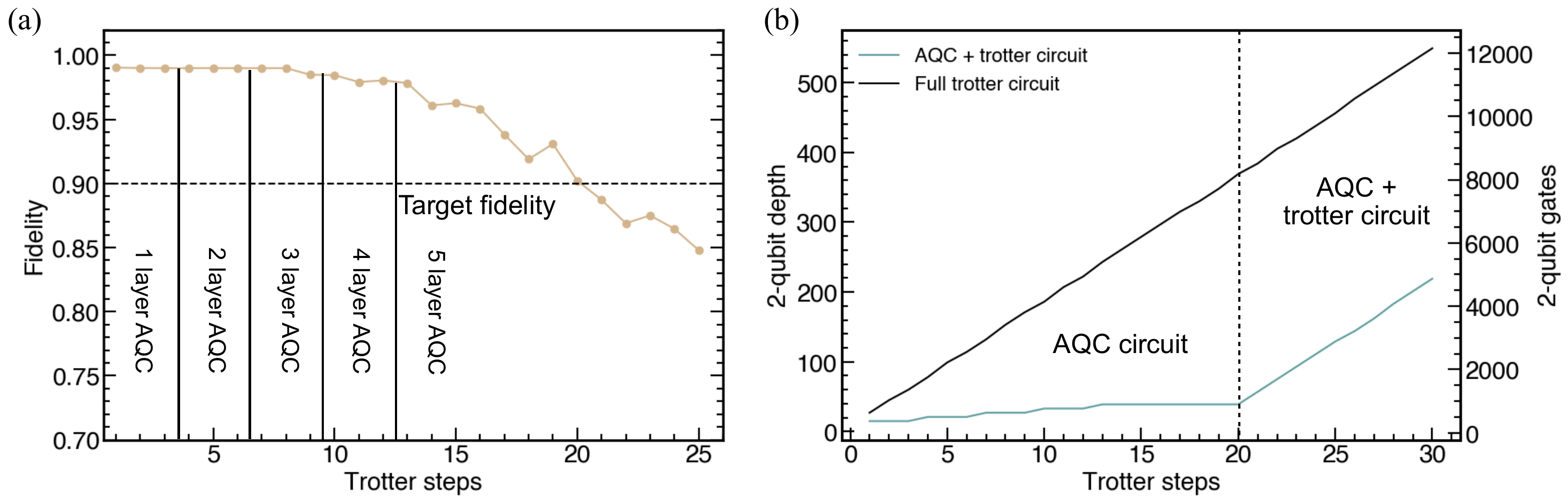}
\caption{\label{aqc_cscocl3_nnn}
\textbf{AQC information of the quantum circuits for CsCoX$_3$ Hamiltonian.}
(a) Fidelity between the original Trotter circuit and the AQC circuit.
The deepest ansatz we employ consists of five Trotter layers combined with the ground-state circuit.
(b) Comparison of the 2-qubit depth and gate count between the original Trotter circuit and the AQC+Trotter circuit. 
After 20 Trotter steps, the depth and gate costs increase linearly as the Trotter circuit is concatenated to perform the simulation.} 
\end{figure*}

For the CsCoX$_3$ Hamiltonian in Equation~\ref{NNN}, we start with the ground-state circuit that has the 2-qubit depth of 12 and 294 two-qubit gates.
Instead of using the entire Trotter step in our ansatz, we only include the circuit involving the NN interaction, which corresponds to the first part of the circuit shown in figure~\ref{trotter_circuit}(c).
This reduces the circuit depth, and the time evolution with small NNN interactions can still be effectively parametrized, as the fidelity shown in figure~\ref{aqc_cscocl3_nnn}(a).
Based on the target fidelity of 90 \%, we perform the AQC circuit for up to 20 Trotter steps, achieving a fidelity of 90.2 \%.
The deepest ansatz we are able to employ is the ground-state circuit with 5 Trotter steps, as the optimization of the ansatz with 6 Trotter steps for the ground-state circuit with 18 Trotter steps could not be completed within the one-day node wall-time limit.
The comparison of the 2-qubit gate count and 2-qubit depth are shown in figure~\ref{aqc_cscocl3_nnn}(b).
The deepest circuit after AQC reaches a 2-qubit depth of 219 and contains 4,908 two-qubit gates.
Compared to the original circuit, the number of two-qubit gates is reduced by 7,245, and the two-qubit depth is reduced by 330.
The significant amount of circuit reduction is due to bypassing the implementation of the time-evolution circuit based on the NNN interaction, which involves swap networks at the early-time period.
The two-qubit gates from the AQC circuit account for 17.8 \% of those in the deepest quantum circuit.

\begin{figure*}[h!]
\centering
\includegraphics[width=0.9\textwidth]{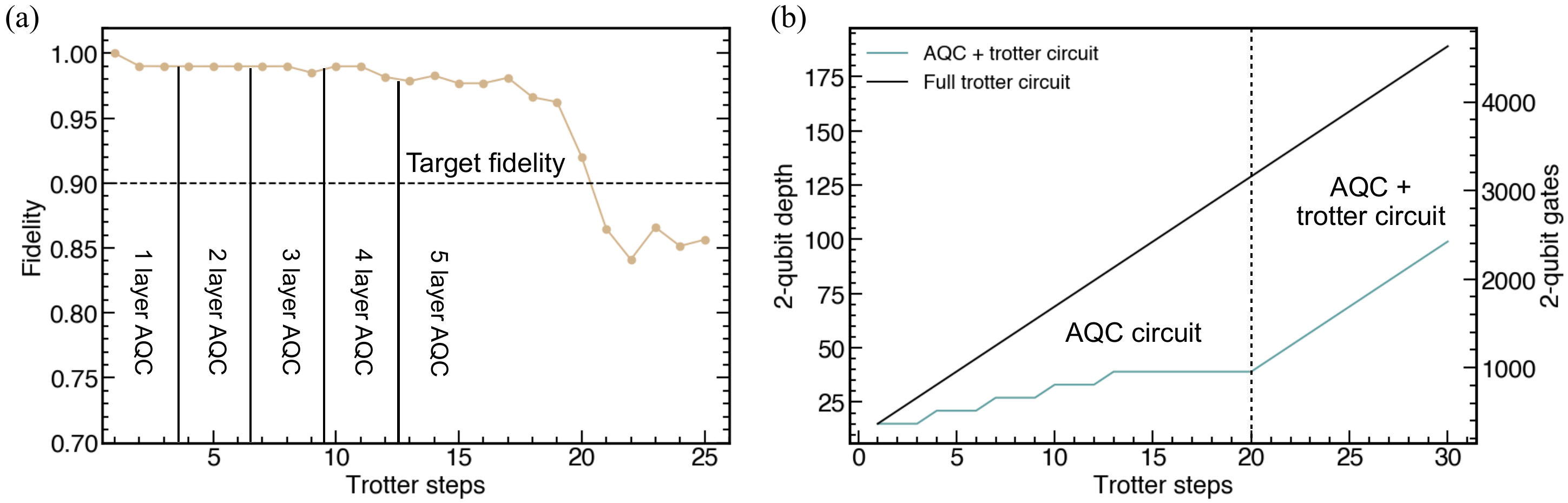}
\caption{\label{aqc_cscocl3_nn}
\textbf{AQC information of the quantum circuits for two-soliton continuum.}
(a) Fidelity between the original Trotter circuit and the AQC circuit.
The deepest ansatz we employ consists of five Trotter layers combined with the ground-state circuit.
(b) Comparison of the 2-qubit depth and gate count between the original Trotter circuit and the AQC+Trotter circuit. 
After 20 Trotter steps, the depth and gate costs increase linearly as the Trotter circuit is concatenated to perform the simulation.} 
\end{figure*}

For the circuit with two-soliton continuum, we start with the same ground-state circuit as in the CsCoX$_3$ Hamiltonian.
For the purpose of comparison, we perform AQC circuit with the same trotter steps in the ansatz circuit as we used in the AQC circuit with NNN circuits, and the corresponding fidelity is shown in figure~\ref{aqc_cscocl3_nn}(a). 
While there is only NN interaction in the circuit, it is expected that the quantum state is less entangled and the corresponding AQC circuit is easier to prepare compared to the previous case with both NN and NNN interaction.
Thus, the fidelity shown in figure~\ref{aqc_cscocl3_nn}(a) is slightly better than the one shown in figure~\ref{aqc_cscocl3_nnn}(a).
Based on the target fidelity, we also prepare the AQC circuit for the original circuit with Trotter steps up to 20. 
The comparison of the 2-qubit gate count and 2-qubit depth are shown in figure~\ref{aqc_cscocl3_nn}(b). 
The deepest circuit after AQC reaches a 2-qubit depth of 99 and contains 2,424 two-qubit gates.
Compared to the original circuit, the number of two-qubit gates is reduced by 2,205, and the two-qubit depth is reduced by 90.
The two-qubit gates from the AQC circuit account for 39.4 \% of those in the deepest quantum circuit.

Notably, for short period of time, the circuit could be optimized based on the light-cone structure of the correlator or the Green's function and further benefit the AQC process. However, for longer time period, as the correlator and Green's function propagate through the entire circuit, the reduction of the quantum gates will be limited.

\subsection{Resolution of the simulation}
In quantum simulation, the resolution of the Fourier-transformed (FT) spectrum is set by the number of qubits and the cost of implementing two-qubit gates, which are the primary source of error.
For FT spectrum, the momentum resolution is \(\Delta k = 2\pi/n\) for \(n\) qubits, while the energy resolution is \(\Delta \omega = 2\pi/(N dt)\), where \(N\) is the number of Trotter steps and \(dt\) the time interval.
The choice of \(dt\) is governed by the Nyquist sampling theorem: to accurately capture excitations up to energy \(E\), the sampling rate must satisfy \(\pi/dt \geq E\). This ensures that the highest frequency component is resolved without aliasing in the FT spectrum. Combining these:
\begin{equation}
\label{resolution}
\Delta k \Delta \omega = \frac{4\pi^2}{N n dt} = \frac{4\pi E}{N n}.
\end{equation}
Under gate constraints, either the total number of two-qubit gates \(G\) or the circuit depth \(D\) is fixed to achieve a target error threshold due to the hardware noise.
For the purpose of analyzing the relationship between gate cost and spectral resolution, we assume that most two-qubit gates arise from the time-evolution block of the circuit in Figure~\ref{overview}.
We do not account for any compression techniques such as AQC, since their effectiveness depends on the specific model and its entanglement structure. 
For a 1D XXZ chain using first-order Trotterization with periodic boundary conditions (PBC), each Trotter step requires $3n$ two-qubit gates, giving a total gate count $G = 3nN$.
With a fixed circuit depth $D$, the maximum number of Trotter steps is $N = D/6$. 
The combined resolution in terms of G and D becomes: 
\begin{align}
\label{resolution_1D}
\Delta k \Delta \omega = \frac{12\pi E}{G}, \quad
\Delta k \Delta \omega = \frac{24\pi E}{D n}.
\end{align}
For a 2D lattice of size $n_x \times n_y$ with PBC, each Trotter step requires \(6n_x n_y\) two-qubit gates, giving \( G = 6n_x n_y N \). 
As a 2D lattice has coordination number four and each two-qubit interaction is implemented using three two-qubit gates, 
the maximum number of Trotter steps is $N = D/12$.
The combined resolution is:
\begin{align}
\label{resolution_2D}
\Delta k_x \Delta k_y \Delta \omega = \frac{48\pi^2 E}{G}, \quad
\Delta k_x \Delta k_y \Delta \omega = \frac{96\pi^2 E}{n_x n_y D}.
\end{align}

\subsection{Convergence of the bond dimension for MPS simulation}
\begin{figure*}
\centering
\includegraphics[width=0.9\textwidth]{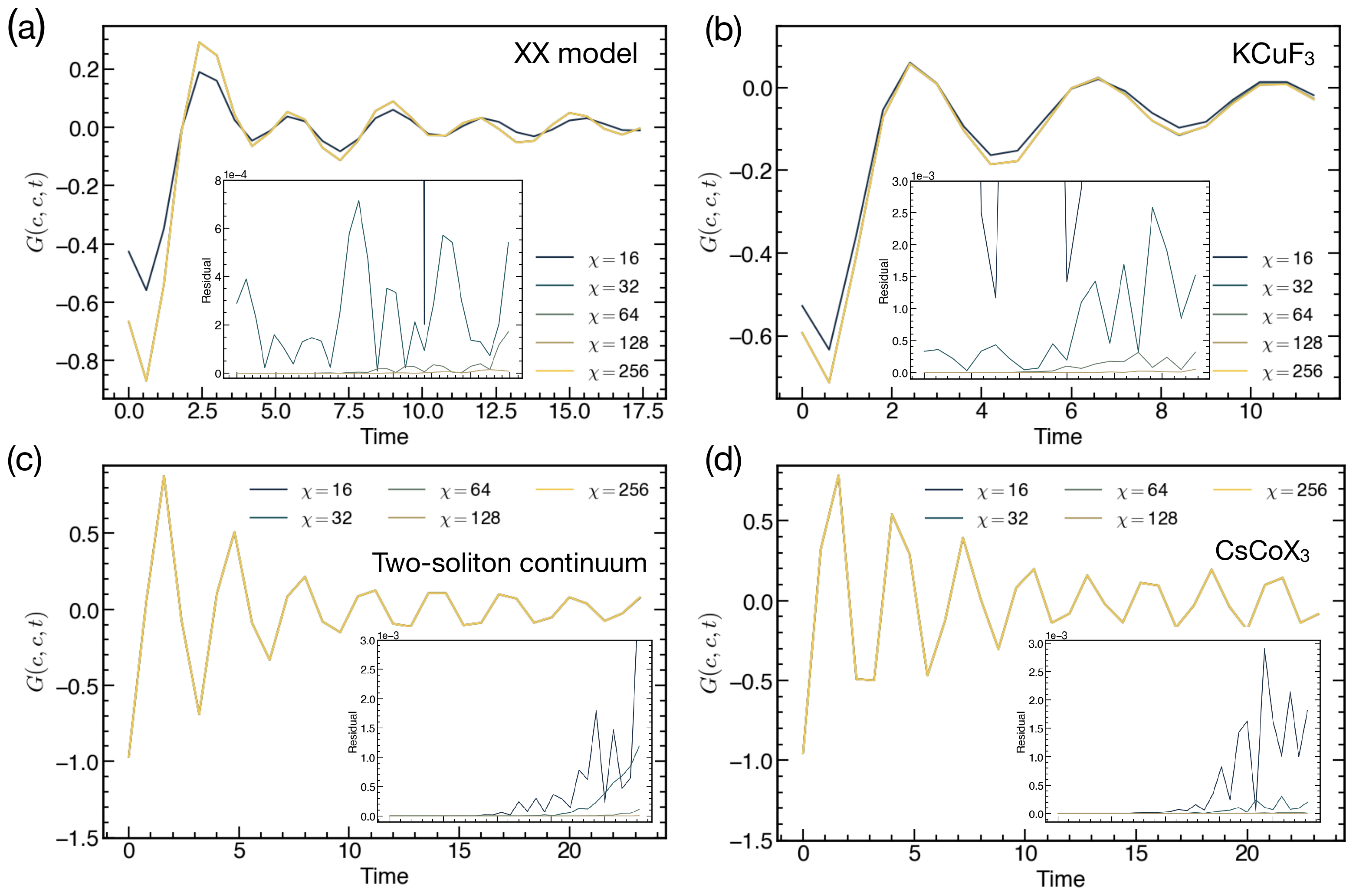}
\caption{\label{bond_dimension}
\textbf{Bond-dimension convergence test} for the observable $G(c,c,t)$ for each model, where $c$ is the center site. 
The inset shows the residual between the observable calculated with different bond dimensions and that with the highest bond dimension, $\chi = 256$.
The results converge to $10^{-4}$ for both the XX model and KCuF$_3$, and to $10^{-5}$ for the CsCoX$_3$ model and two-soliton continuum.
} 
\end{figure*}

We perform an MPS-based simulation and use it as one of the benchmarks for our quantum simulation, as it can handle 1D systems and serves as a noiseless reference.
To represent the wavefunction as a product of matrices, the bond dimension, $\chi$, serves as the primary parameter controlling the capacity to capture many-body entanglement.
Here, we perform a convergence test of the bond dimension for the RDF $G(c,c,t)$, where $c$ refers to the central site, using bond dimensions ranging from 16 to 256 for each model in this research.
Based on the convergence results and the residuals of each bond dimension relative to $\chi = 256$ (Figure. \ref{bond_dimension}), we use $\chi = 128$ as the benchmark in this work. 
The residuals converge to $10^{-4}$ for the XX model and KCuF$_3$, and to $10^{-5}$ for the models with NNN interaction.

\subsection{Approximated error}
In this work, we employ several approximations to evaluate the spectrum, including ground-state preparation (GSP) using a variational ansatz and an AQC circuit. In addition, the results are affected by errors originating from the quantum hardware. 
In this section, we analyze the impact of these different error sources by taking the DMRG ground state with full Trotterized time evolution as a reference and quantifying deviations via the structural similarity index (SSIM) of the measured RGF.
We evaluated the RGF and the corresponding spectra for the GSP and GSP+AQC protocols using MPS simulations, which serve as noiseless benchmarks.
In addition, we evaluate the RGF and the spectrum on \textit{ibm\_kingston}, \textit{ibm\_pittsburgh}, and \textit{ibm\_boston} to examine how device quality affects the results, although it is straightforward that higher-quality devices yield better performance.

\begin{figure*}
\centering
\includegraphics[width=1\textwidth]{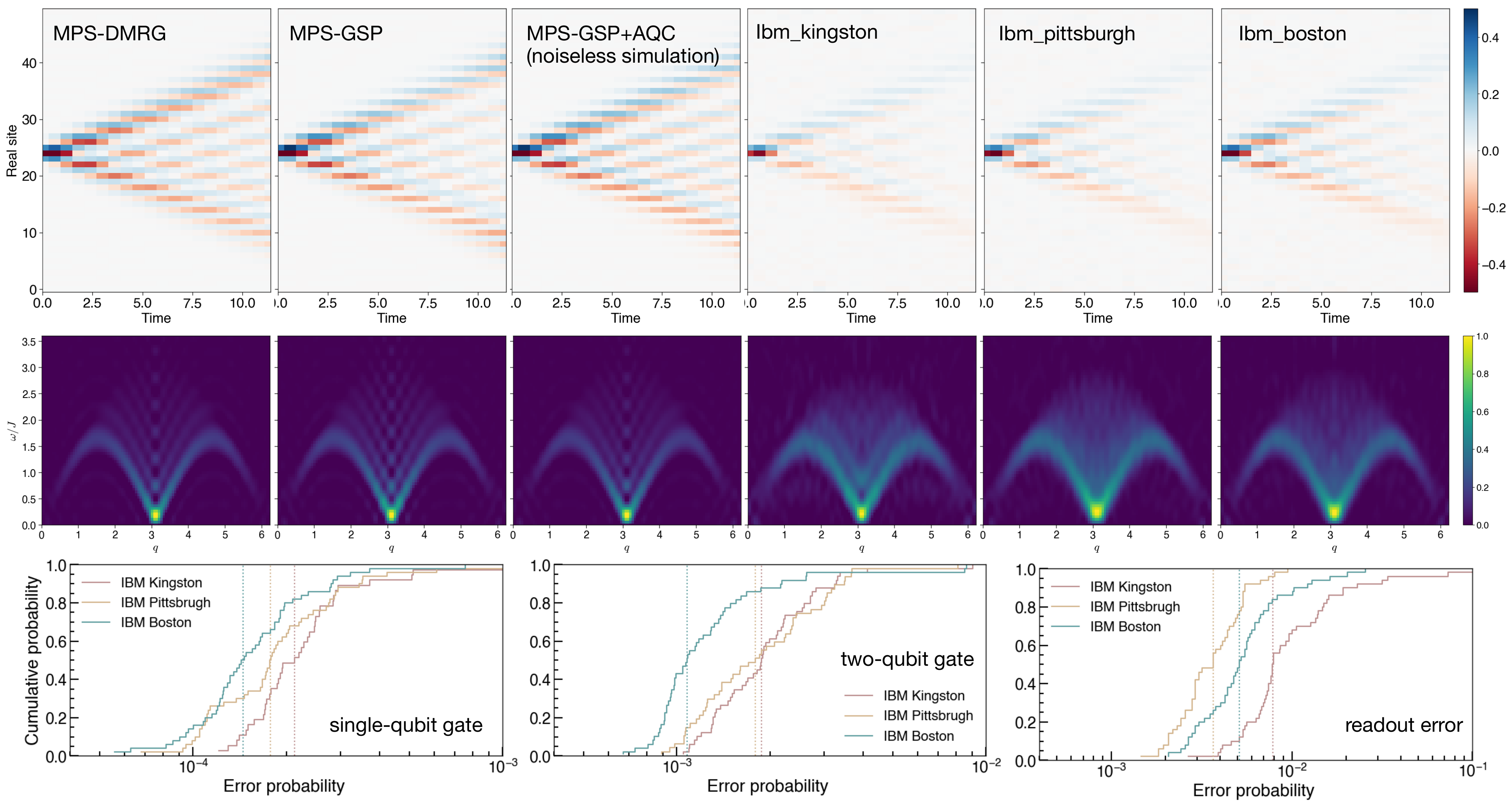}
\caption{\textbf{Comparison of quantum simulated KCuF$_3$ spectrum with noiseless simulations.}
The fidelity of the GSP and AQC is 83 \% and 90 \% respectively. 
The median error rate of single-qubit gate are $2.130 \times 10^{-4}$, $ 1.778 \times 10^{-4}$, and $1.449 \times 10^{-4}$; those of two-qubit gate are $1.877 \times 10^{-3}$, $1.792 \times 10^{-3}$, and $1.080 \times 10^{-3}$; and those of readout are $7.876 \times 10^{-3}$, $3.662 \times 10^{-3}$, and $5.123\times 10^{-3}$, for $ibm\_kingston$ on August 25$^{th}$, 2025, $ibm\_pittsburgh$ on Octorber 13$^{th}$, 2025, and $ibm\_boston$ on December 22$^{nd}$, 2025, respectively.}
\label{fig:error_kcuf3}
\end{figure*}

We consider KCuF$_3$ as a representative case.
The spectral features are well captured by the GSP+AQC approximation, as shown in figure~\ref{fig:error_kcuf3}, with the corresponding SSIM exceeding 99\%, as summarized in table~\ref{tab:error}.
Notably, the relative SSIM between Green’s functions and spectra reflects the different sensitivity of time- and frequency-domain observables to phase coherence, dispersion, and spectral-weight redistribution.
The improvement of the device yields a better spectrum, as shown in figure~\ref{fig:error_kcuf3}
From $ibm\_kingston$ (Heron r2 device) to $ibm\_boston$ (best Heron r3 device), we observe an improved capture of the overall RGF and enhanced spectrum quality, with the corresponding SSIM approaching the noiseless result (MPS-GSP+AQC).

\begin{table}
\centering
\begin{tabular}{l| c | c |c |c |c| c}
\textrm{Approximation}&
\textrm{MPS-GSP}&
\textrm{MPS-GSP+AQC} & 
\textrm{$ibm\_kington$} & 
\textrm{$ibm\_pittsburgh$} & 
\textrm{$ibm\_boston$} \\
\hline
KCuF$_3$ - RGF & 0.892 & 0.859 & 0.626 & 0.658 & 0.705  \\
KCuF$_3$ - spectrum & 0.990 & 0.990 & 0.845 & 0.877 & 0.900 \\
\end{tabular}
\caption{\label{tab:error}
SSIM comparison of the RGF and spectrum against the MPS result using DMRG state for KCuF$_3$ (Figs. \ref{fig:error_kcuf3}).
}
\end{table}

\subsection{Metrics}

The Mean Square Error (MSE) provides the most direct measure of the quality of the predictor $\hat{y}$ relative to the baseline $y$, where the target spectrum $y_{target}$ is compared with the experimental spectrum $y_{expt}$.
MSE is defined as
\begin{equation}
\text{MSE} = \frac{1}{n} \sum_i^n (y_i- \hat{y}_i)^2 = \frac{(y_{target}-y_{expt})^2} {n}
\end{equation}
where $n$ denotes the number of predictions, corresponding to the total number of pixels in the spectrum.

To quantify the differences between spectra, we evaluate the Wasserstein distance (also known as the Earth Mover’s Distance~\cite{rubner1998metric}), which measures the minimum cost required to transform one intensity distribution into another. 
The Wasserstein distance is sensitive to shifts and broadening of spectral features, making it well suited for comparing dispersive structures in dynamical spectra.
The Wasserstein distance between x and y is calculated using the \textsc{Scipy}~\cite{2020SciPy-NMeth} package, according to the formula
\begin{equation}
l_1({u, v}) = {\inf}_{\pi \in \Gamma(u, v)} \int ||(x-y)||_2 d\pi (x,y)
\end{equation}
where $\Gamma(u,v)$ is the set of all transport plans with marginals $u$ and $v$.

Structural Similarity Index Measure (SSIM)~\cite{wang2004image} is a widely used metric for comparing image similarity, particularly in computer science and image processing. The SSIM index is computed between two windows of pixel values $x$ and $y$ of common size, taken from corresponding locations in two images. These SSIM values can be aggregated across the full image by averaging or other variations.
In a simple special case, the SSIM measure between $x$ and $y$ is given by:
\begin{equation}
\text{SSIM}(x,y) = \frac{(2\mu_x \mu_y + c_1)(2\sigma_{xy} + c_2)}{(\mu_x^2 + \mu_y^2 + c_1)(\sigma_x^2 + \sigma_y^2 + c_2)}
\end{equation}
where:
$\mu_x$ ($\mu_y$) and $\sigma_x^2$ ($\sigma_y^2$) denote the sample means and variances of $x$ ($y$), respectively. $\sigma_{xy}$ denotes the sample covariance between $x$ and $y$. $c_1$ and $c_2$ are small constants introduced to stabilize the division, where $c_1$ and $c_2$ are set to the standard values $0.01^2$ and $0.03^2$ based on the data range of the normalized spectrum equals 1~\cite{wang2004image}.
The SSIM is evaluated using the~\textsc{scikit-image} package~\cite{van2014scikit} with window size of 11 by 11.

To further assess the spectral characteristics, we evaluate features such as the spectral weight and the positions of the main peaks, providing a more detailed benchmark of the quantum simulations.
Furthermore, when the spectrum exhibits a discernible distribution, which may reflect specific quasiparticle dynamics, it is worthwhile to evaluate the corresponding properties to investigate transport behavior.
To determine the peak position and the associated spectral weight, we fit the line-scan data using a Gaussian function implemented in~\textsc{Lmfit}~\cite{newville2016lmfit}.
Once the peak position is identified, we integrate the spectrum over a window of width $0.5 J$ centered on the peak to account for broadening effects.

For entanglement witness, we employ the normalized form of QFI (nQFI) \cite{scheie2025tutorial}, based on the energy integration of the DSF, defined as 
\begin{equation}
\label{eq:nQFI}
nQFI[q,T] = \frac{1}{S^2} \int^\infty_0 d(\hbar \omega) [\tanh(\frac{\hbar \omega}{2k_BT}) 
(1- e^{-\hbar \omega / k_BT})S_{\alpha,\alpha}(q, \omega)]
\end{equation}
where $S = 1/2$, $T$ is the temperature, $\hbar \omega$ is the energy transfer, 
and  $k_B$ denotes the Boltzmann constant. 
The $nQFI > m$ indicated the system with at least $m+1$-partite entanglement.
We compute Eq.~\ref{eq:nQFI} after renormalizing the total spectrum based on the sum rule as follows.
\begin{equation}
\label{eq:sum}
\frac{\int^\infty_{-\infty} d\omega \int_{BZ} dq \sum_\alpha S_{\alpha, \alpha}(q, \omega)}{\int_{BZ} dq } = S(S+1)
\end{equation}
Notably, a small nQFI does not necessarily imply a low degree of entanglement within the wave function, as it serves as a lower bound for the information about multipartite entanglement. 
Furthermore, we evaluate the two-tangle, $\tau_2$, which quantifies pairwise entanglement as the sum of squared concurrences, $\tau_2 = \sum_{r \neq 0} C_r^2$~\cite{scheie2021witnessing, scheie2025tutorial}. This measure involves the spin-spin correlation function, $C_{zz} = \langle S^z_i S^z_{i+r} \rangle $, which can be obtained via the Fourier transform of the energy-integrated spectrum.
For the isotropic point, the concurrences can be simplified as~\cite{scheie2021witnessing}
\begin{align}
\label{eq:cr}
C_r = 2 \max \{ 0, 2 |{C_{zz}}| - |\frac{1}{4} + C_{zz}| \}
\end{align}
with $\tau_2 = 2\sum_{r\neq0}C_r^2$.
We note that $\tau_2$ is sensitive to experimental artifacts and therefore should be used only as an auxiliary or semi-quantitative metric for entanglement witnessing.

\subsection{Noisy simulation}
To evaluate the effect of gate errors on the quality of the INS spectrum, we employ a global depolarization channel to describe the noise of two-qubit gates, which are most dominant source of noise in our experiments. 
\begin{equation}
\label{depo_channel}
\mathcal{D}(\rho) = (1-p)\rho + p\frac{I}{2^n}
\end{equation}
where $\rho$ is the density matrix and $p$ is the depolarization probability.
To incorporate the error rates from each two-qubit gate into equation~\ref{depo_channel}, we approximate $(1-p)$ as $\prod_i^N (1-p_i)$, where $p_i$ is the gate error and $N$ is the total number of two-qubit gates.

We first take the compiled AQC circuit and evaluate the number of two-qubit gates within the backward light cone from the final measurement at each time step, which serves as $N$ for approximating the total $(1-p)$ factor. Next, to assign an error to each two-qubit gate, we sample it from a Gaussian distribution with the given error rate as the mean. Then, we sample bitstrings from the MPS simulation of the quantum circuit to create the noiseless bitstring dictionary and randomly choose the sampled bitstrings from the dictionary based on the probability of $(1-p)$.
To mimic the maximally mixed state (second term in Equation~\ref{depo_channel}), we sample random bitstrings with probability $p$.
In total, we sample 128000 bitstrings probabilistically, either from the MPS distribution or random bitstrings, with $(1-p)$ informing the probability of encountering noise. The same number of shots was used in the experimental setup. The RGF and the corresponding DSF were computed from the acquired distribution. 

\subsection{Spectrum Broadening}
For experimental measurements, spectral features broaden at finite-temperature because thermal fluctuations increase the number of accessible transitions and reduce quasiparticle lifetimes. In the quantum simulation results, noise generally damps the measured observables, causing a broadening of the signal.

\begin{figure*}
\centering
\includegraphics[width=1\textwidth]{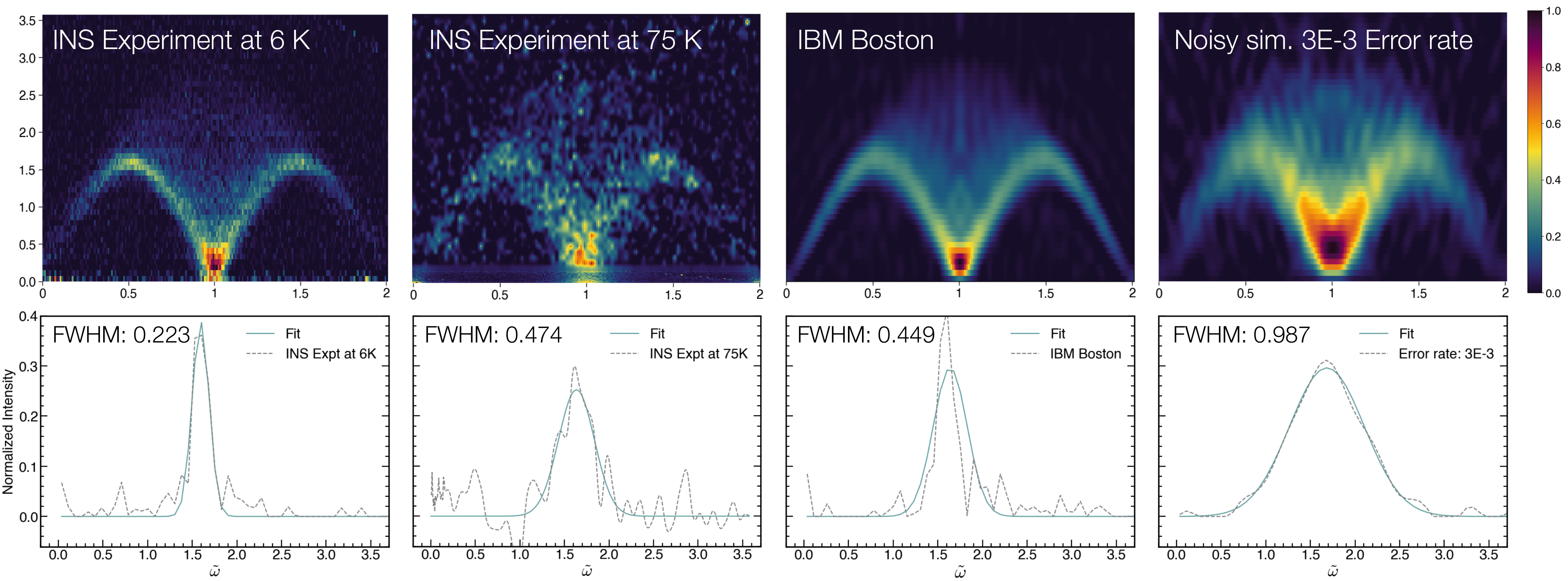}
\caption{\textbf{Comparison of the KCuF$_3$ spectrum under finite-temperature conditions and depolarization noise.}
The spectrum at 75K is replotted based on the data in Ref.~\cite{scheie2022quantum}. 
The full width at half maximum increases as the measurement temperature and error rate increase.
The line scan is measured at the $\pi/2$ point.
}
\label{compare_fwhm}
\end{figure*}

Consistent with these effects, the full width at half maximum (FWHM) of the KCuF$_3$ spectrum at the $\pi/2$ point is larger at higher temperatures and higher error rates, as shown in Figure~\ref{compare_fwhm}. Because quantum noise effectively increases the space of measurement outcomes, it can be interpreted as mimicking certain finite‑temperature effects, suggesting the possibility of defining an effective temperature associated with the noise. However, establishing such a correspondence lies beyond the scope of the present work.

\subsubsection*{Hardware information}
\begin{figure*}
\centering
\includegraphics[width=0.9\textwidth]{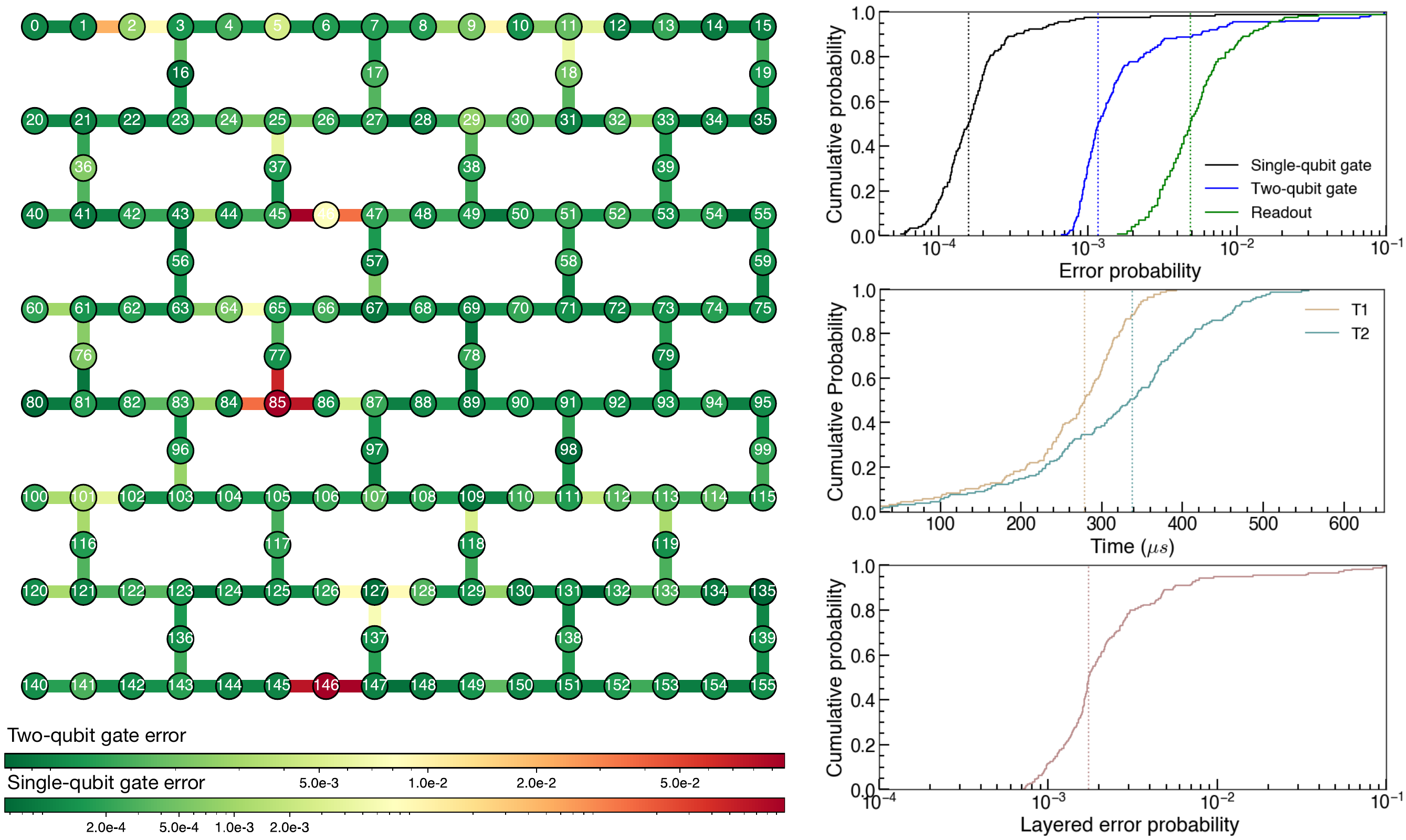}
\caption{Properties of $ibm\_boston$ on December 22$^{nd}$, 2025, for the XX model and KCuF$_3$ experiment.
The single- and two-qubit gate errors across the entire device map are shown in the left panel.
The right panel shows the cumulative percentages of single-qubit errors, two-qubit errors, readout errors, $T_1$, $T_2$, and the layered error from the LF experiment.}
\label{fig:hardware}
\end{figure*}

\begin{figure*}
\centering
\includegraphics[width=0.9\textwidth]{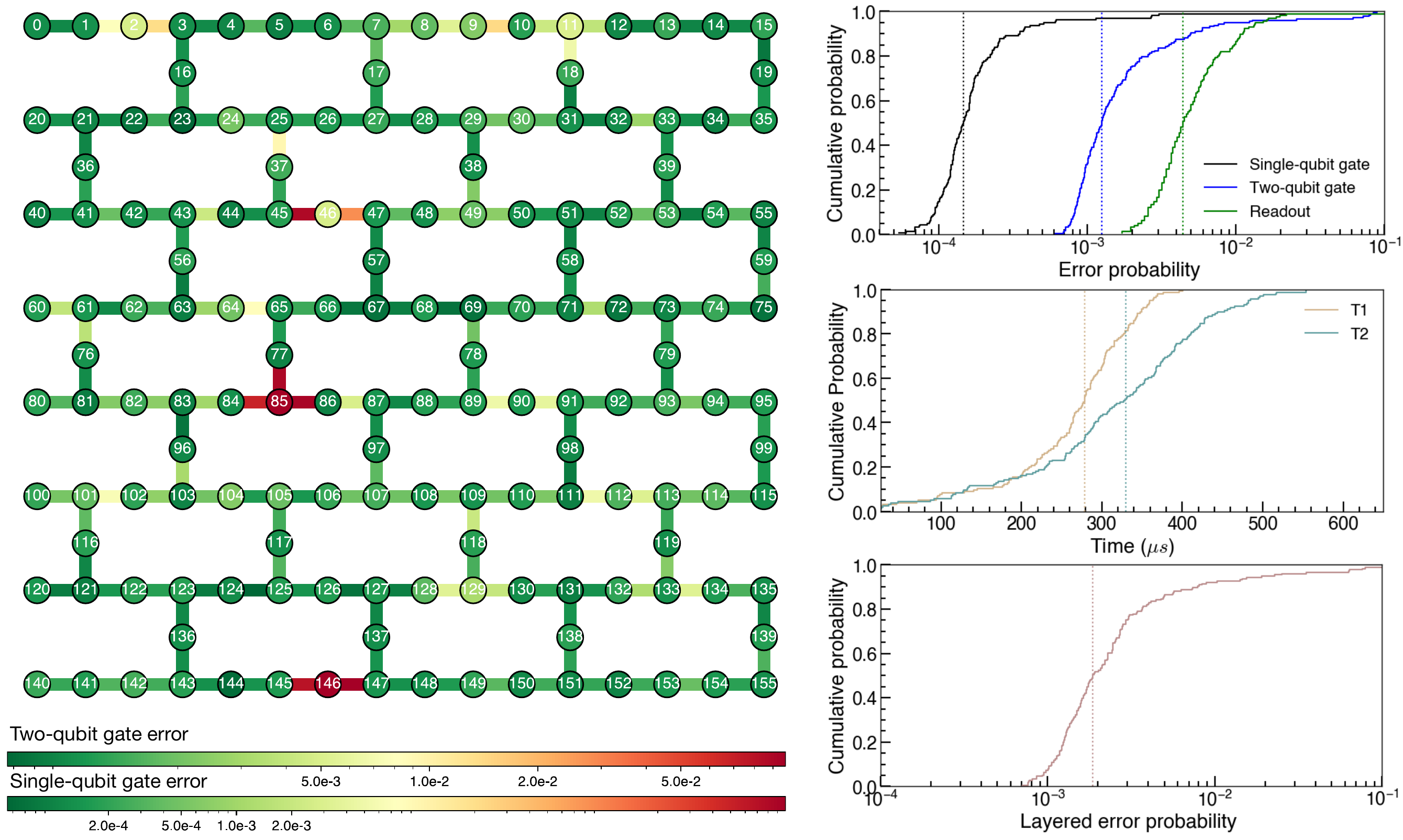}
\caption{Properties of $ibm\_boston$ on January 26$^{th}$, 2026, ibm for the two-soliton continuum and CsCoX$_3$ experiment.
The single- and two-qubit gate errors across the entire device map are shown in the left panel.
The right panel shows the cumulative percentages of single-qubit errors, two-qubit errors, readout errors, $T_1$, $T_2$, and the layered error from the LF experiment.}
\label{fig:hardware_2}
\end{figure*}

The quantum experiments are performed on the IBM Heron r3 processor, $ibm\_boston$, which comprises 156 fixed-frequency transmon qubits \cite{koch2007charge} arranged in a heavy-hex connectivity.
We conduct our experiments in two sets: one involves the XX model and KCuF$_3$, and the other involves the two-soliton continuum and CsCoX$_3$.
The device properties for conducting the experiments for the XX model and KCuF$_3$ are shown in Figure~\ref{fig:hardware}.
The median single-qubit gate, two-qubit gate, and readout error rates are $1.584 \times 10^{-4}$, $1.177 \times 10^{-3}$, and $4.883 \times 10^{-3}$, respectively.
The median $T_1$ and $T_2$ times are  278.65 and 337.76 $\mu s$, respectively.
On the other hand, the device properties of the experiments for the two-soliton continuum and CsCoX$_3$ are shown in figure~\ref{fig:hardware_2}.
The median single-qubit gate, two-qubit gate, and readout error rates are $1.471 \times 10^{-4}$, $1.256 \times 10^{-3}$, and $4.395 \times 10^{-3}$, respectively.
The median $T_1$ and $T_2$ time are 278.67 and 329.76 $\mu s$, respectively.
To further benchmark the device quality and optimize the layout selection, we perform a layer fidelity (LF) experiment via randomized benchmarking (RB) \cite{mckay2023benchmarking}, yielding median error rates of $1.734 \times 10^{-3}$ and $1.867 \times 10^{-3}$ for the two sets of experiments, as shown in the bottom-right panels of figure~\ref{fig:hardware} and figure~\ref{fig:hardware_2}, respectively.

\begin{figure*}
\centering
\includegraphics[width=0.9\textwidth]{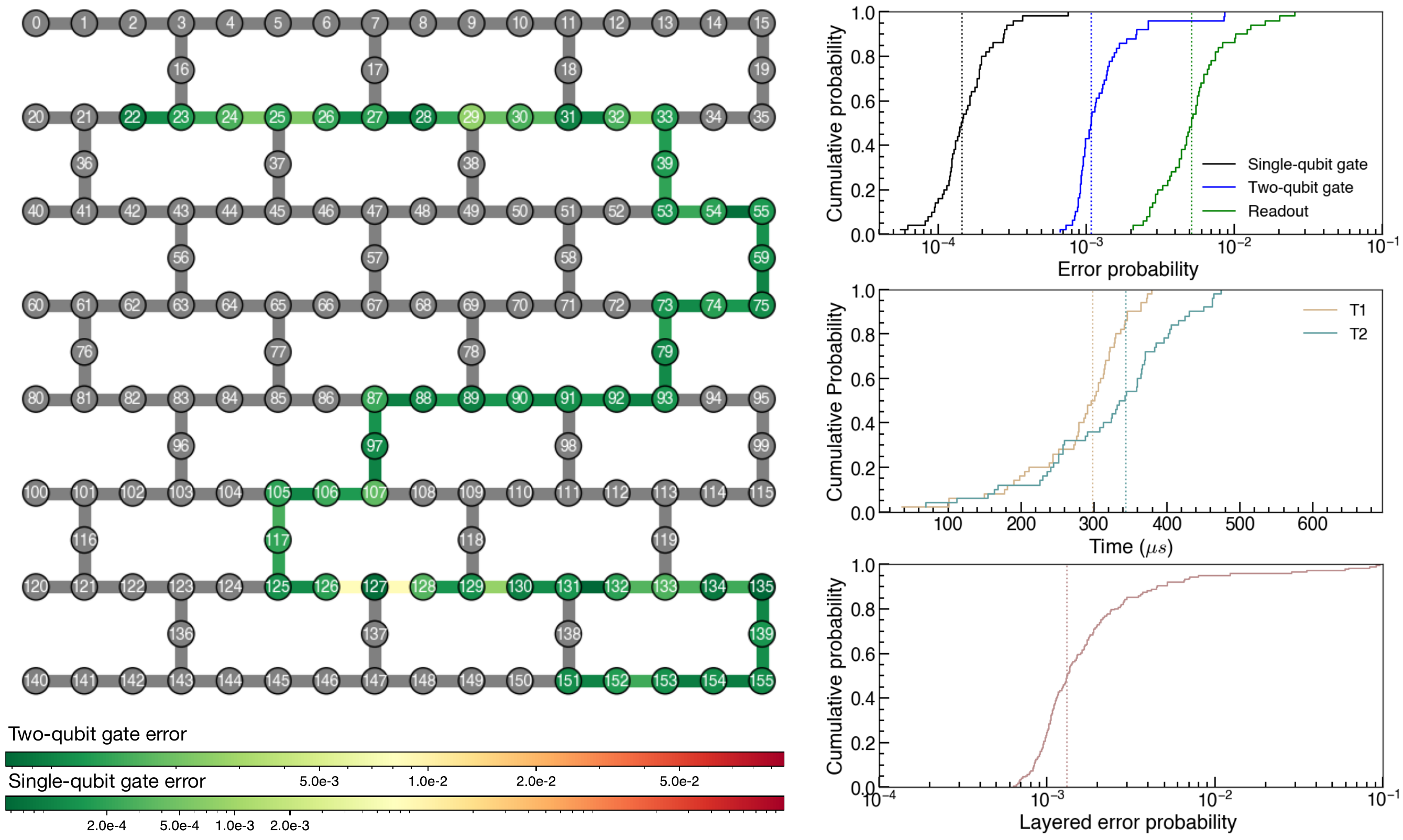}
\caption{Properties of chosen 50 qubits on $ibm\_boston$ on December 22$^{nd}$, 2025, for the XX model and KCuF$_3$ experiment.
The layered error evaluated in the LF experiment with the layered connectivity of the quantum circuit used in this research.}
\label{fig:hardware_layout}
\end{figure*}

\begin{figure*}
\centering
\includegraphics[width=0.9\textwidth]{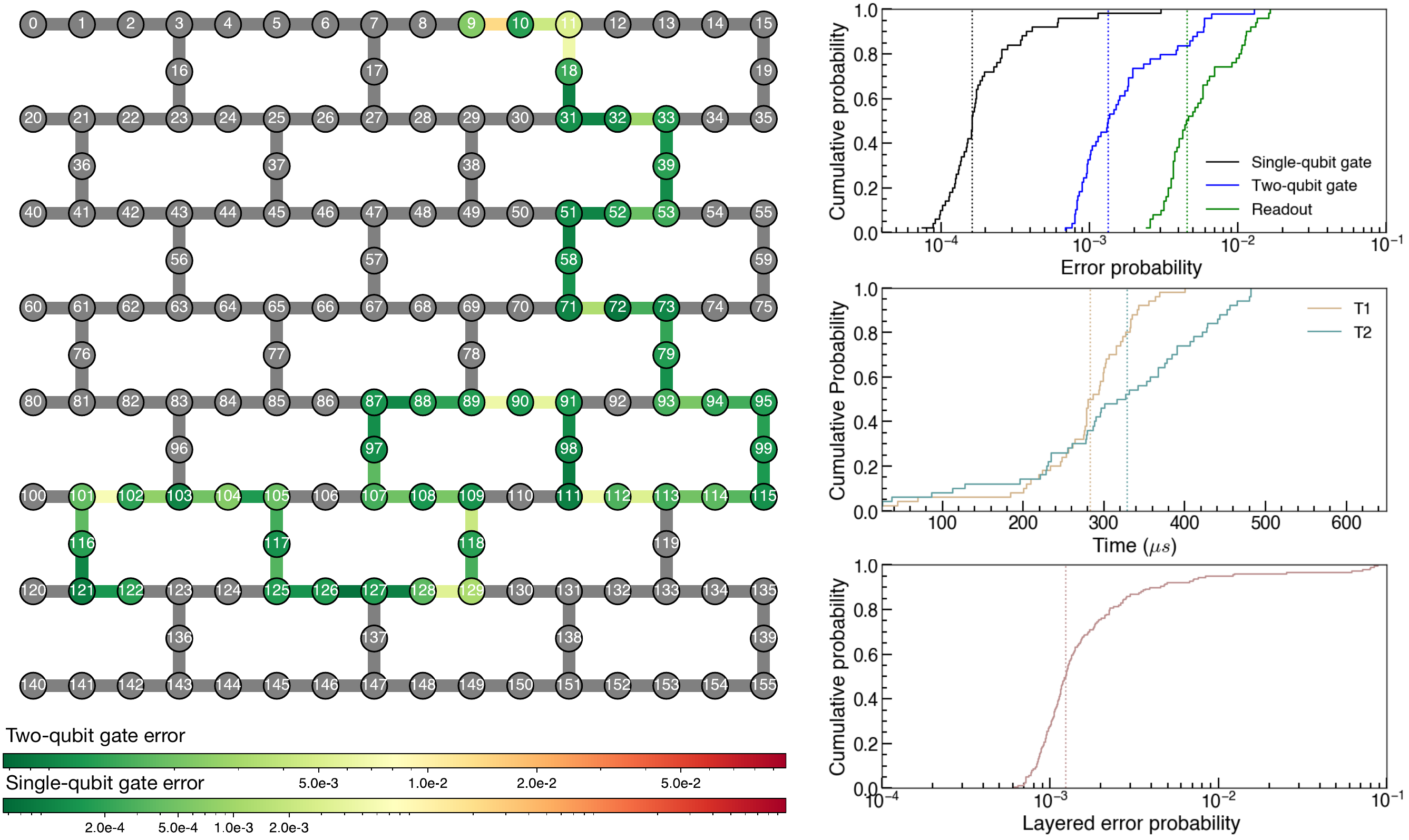}
\caption{Properties of chosen 50 qubits on $ibm\_boston$ on January 26$^{th}$, 2026, for the two-soliton continuum and CsCoX$_3$ experiment.
The layered error evaluated in the LF experiment with the layered connectivity of the quantum circuit used in this research.}
\label{fig:hardware_layout_2}
\end{figure*}

To choose the optimal layout for our 50-qubit experiment, we first randomly sample 10,000 candidate layouts generated by \texttt{mapomatic} \cite{PRXQuantum.4.010327}.
We then determine the optimal layout based on the fidelity evaluated from the layered error rates and the readout error rates from the device. 
The layout properties of the experiments for XX model and KCuF$_3$ are shown in figure~\ref{fig:hardware_layout}.
The median single-qubit gate, two-qubit gate, and readout error rates are $1.450 \times 10^{-4}$, $1.080 \times 10^{-3}$, and $5.127 \times 10^{-3}$, respectively.
The median $T_1$ and $T_2$ time are 298.40 and 343.75 $\mu s$, respectively.
The layout properties of the experiments for two-soliton continuum and CsCoX$_3$ are shown in figure~\ref{fig:hardware_layout_2}.
The median single-qubit gate, two-qubit gate, and readout error rates are $1.632 \times 10^{-4}$, $1.342 \times 10^{-3}$, and $4.578 \times 10^{-3}$, respectively.
The median $T_1$ and $T_2$ time are 283.59 and 329.22 $\mu s$, respectively.
The LF experiment is implemented using the layered connectivity of the quantum circuits employed in this research, with a median error rate of $1.316 \times 10^{-3}$ and $1.234 \times 10^{-3}$, as shown in the bottom-right panels of figure~\ref{fig:hardware_layout} and figure~\ref{fig:hardware_layout_2}, respectively.


\begin{table}
\begin{tabular}{cccccccc} 
Quantum experiments & Median & \shortstack{T1 ($\mu s$)} & \shortstack{T2 ($\mu s$)}  & 
\shortstack{1Q ($10^{-4}$)}&
\shortstack{2Q  ($10^{-3}$)} & 
\shortstack{RO  ($10^{-3}$)} & 
\shortstack{LF  ($10^{-3}$)} \\
\hline
XX model / KCuF$_3$ & overall   &  278.65 &  337.76  &  1.584  &  1.177  &  4.883 & 1.734 \\ \hline
XX model / KCuF$_3$ & layout   &  298.40 &  343.75  &  1.450 &  1.080  &  5.127 & 1.316 \\ \hline

2-soliton / CsCoX$_3$ & overall  &  278.67 &  329.76 &  1.471  &  1.256  &  4.395 & 1.867\\ \hline
2-soliton / CsCoX$_3$ & layout  &  283.59 &  329.22 &  1.632 &  1.342  &  4.578 & 1.234 \\ \hline
\end{tabular}
\caption{\label{table: Neel cali}
Calibration data for the 50-qubit quantum simulation for XX model/KCuF$_3$ and two-soliton/CsCoX$_3$.}
\end{table}

\end{document}